\documentclass[11pt,a4]{article}
\usepackage{latexsym,amssymb} 
\newcommand{\be}{\begin{equation}}
\newcommand{\ee}{\end{equation}}
\newcommand{\beqs}{\begin{eqnarray}}
\newcommand{\eeqs}{\end{eqnarray}}
\def\vv{{\cal V}}
\def\({\left(}
\def\){\right)}

\newcommand{\Exc}[1]{{${\rm E}_{{#1}({#1})}$}}

\def\mxth{\mathsurround=0pt }
\def\xversim#1#2{\lower2.pt\vbox{\baselineskip0pt \lineskip-.5pt
x  \ialign{$\mxth#1\hfil##\hfil$\crcr#2\crcr\sim\crcr}}}

\renewcommand{\a}{\alpha}
\renewcommand{\b}{\beta}

\renewcommand{\d}{\delta}
\newcommand{\pa}{\partial}
\newcommand{\g}{\gamma}

\newcommand{\m}{\mu}
\newcommand{\n}{\nu}

\newcommand{\nn}{\nonumber}

\def\be{\begin{equation}}
\def\ee{\end{equation}}
\def\bea{\begin{eqnarray}}
\def\eea{\end{eqnarray}}

\newcommand{\ft}[2]{{\textstyle\frac{#1}{#2}}}

\newcommand{\eqn}[1]{(\ref{#1})}
\def\bfone{\relax{\rm 1\kern-.35em 1}}

\begin{document}
\begin{titlepage}

\begin{flushright}\small 
ITP-UU-07/26 \\ SPIN-07/17 \\ ENSL-00146707
\end{flushright}
%
\vskip 10mm
\begin{center}
{\Large {\bf THE MAXIMAL D$=$4 SUPERGRAVITIES}}
\end{center}
\vskip 8mm

\begin{center}
{{\bf Bernard de Wit}\\
 Institute for Theoretical Physics \,\&\, Spinoza Institute,\\  
Utrecht University, Postbus 80.195, NL-3508 TD Utrecht, 
The Netherlands\\
{\tt b.dewit@phys.uu.nl}}
\vskip 2mm
{{\bf Henning Samtleben}\\
Laboratoire de Physique, ENS Lyon,\\ 
46 all\'ee d'Italie, F-69364 Lyon CEDEX 07, France \\
{\tt henning.samtleben@ens-lyon.fr}}
\vskip 2mm
{{\bf Mario Trigiante}\\
Dept. of Physics, Politecnico di Torino,\\
Corso Duca degli Abruzzi 24, I-10129 Torino, Italy \\
{\tt mario.trigiante@to.infn.it}}
\vskip 4mm

\end{center}

\vskip .2in

\begin{center} {\bf Abstract } \end{center}
\begin{quotation}\noindent
  All maximal supergravities in four space-time dimensions are
  presented. The ungauged Lagrangians can be encoded in an ${\rm
    E}_{7(7)} \backslash {\rm Sp}(56;\mathbb{R})/{\rm GL}(28)$ matrix
  associated with the freedom of performing electric/magnetic duality
  transformations. The gauging is defined in terms of an embedding
  tensor $\Theta$ which encodes the subgroup of \Exc7 that is realized
  as a local invariance. This embedding tensor may imply the presence
  of magnetic charges which require corresponding dual gauge fields.
  The latter can be incorporated by using a recently proposed
  formulation that involves tensor gauge fields in the adjoint
  representation of \Exc7.  In this formulation the results take a
  universal form irrespective of the electric/magnetic duality basis.
  We present the general class of supersymmetric and gauge invariant
  Lagrangians and discuss a number of applications.
\end{quotation}
\end{titlepage}
\eject  
\section{Introduction}
Maximal supergravity in four space-time dimensions contains 28 vector
gauge fields, which, in principle, can couple to charges assigned to
the various fields. To preserve supersymmetry these gauge field
interactions must be accompanied by masslike terms for the fermions
and a scalar potential, as was first exhibited in the gauging of ${\rm
  SO}(8)$ \cite{deWitNic}. In general it is far from obvious which
gauge groups are admissible and will lead to a supersymmetric
deformation of the ungauged Lagrangian. Initially non-compact versions
and/or contractions of ${\rm SO}(8)$ were shown to also lead to viable
gaugings \cite{Hull}, followed, much more recently, by the so-called
`flat' gauge groups \cite{AndDauFerrLle} that one obtains upon
Scherk-Schwarz reductions \cite{ScherkSchwarz,CSS} of
higher-dimensional theories, as well as by several other
non-semisimple groups \cite{Hull2}.

In \cite{dWST1} we presented an {\it ab initio} analysis of all
possible gaugings of four-dimensional maximal supergravity (this was
reviewed in \cite{dWST2}). The gauge group, ${\rm G}_g$, is a subgroup
of the \Exc7 duality group that leaves the combined field equations
and Bianchi identities invariant. The decomposition of the gauge group
generators in terms of the \Exc7 generators is parametrized by the
so-called embedding tensor $\Theta$, which determines not only the
gauge-covariant derivatives, but also the so-called $T$-tensors that
define the masslike terms and the scalar potential. The admissible
embedding tensors can be characterized group-theoretically and this
enables a systematic discussion of all possible gaugings. In
\cite{dWST1}, several examples were presented which demonstrate how
one can conveniently analyze various gaugings in this way. Another
example, which is relevant for IIB flux compactifications, was worked
out in \cite{dWST3}.  The same strategy has been applied to
maximal supergravity in various space-time 
dimensions~\cite{NicSam,dWST5,SamtlebenWeidner,SamtlebenWeidner2},
as well as to theories with a lower number of
supercharges~\cite{Schon:2006kz,dVdW}. 

In this paper we present a complete analysis of all gaugings of
maximally supersymmetric four-dimensional supergravity. We establish
that a gauging is in fact completely characterized by the embedding
tensor, which is subject to two constraints. One constraint, which is
linear, follows from supersymmetry and implies that the embedding
tensor belongs to the ${\bf 912}$ representation of \Exc7. A second
constraint is quadratic and implies that the square of the embedding
tensor does not contain the ${\bf 133} + {\bf 8645}$ representation.
This constraint ensures the closure of the gauge group. Furthermore it
implies that the embedding tensor is gauge invariant, and it ensures
that the charges can always be chosen in the electric subsector upon a
suitable electric/magnetic duality transformation. In this approach
one can establish the consistency of the gauging prior to evaluating
the explicit Lagrangian. Any given embedding tensor that satisfies
these two constraints, defines a consistent supersymmetric and gauge
invariant Lagrangian. In fact, we will present universal expressions
for the Lagrangian and the supersymmetry transformations of gauged
$N=8$ supergravities, encoded in terms of the embedding tensor. The
fermionic masslike terms and the scalar potential have a unique
structure in terms of the so-called $T$-tensor, which is linearly
proportional to the embedding tensor. Here we should perhaps emphasize
that we our results are obtained entirely in a four-dimensional setting.
As is well known, gaugings can originate from the compactification of
a higher-dimensional theory with or without fluxes, or from
Scherk-Schwarz reductions. But whatever their origin, the
four-dimensional truncations belong to the class of theories discussed
in this paper, provided they are maximally supersymmetric
(irrespective of whether the theory will have maximally supersymmetric
groundstates).

A gauging can involve both magnetic and electric charges, each of
which will require corresponding gauge fields. These can be
accommodated by making use of a new formalism \cite{dWST6}, which, in
the case at hand, requires the presence of tensor gauge fields
transforming in the (adjoint) ${\bf 133}$ representation of \Exc7.
Neither the magnetic vector fields nor the tensor fields lead to
additional degrees of freedom owing to the presence of extra gauge
invariances associated with these fields. Because of the extra fields,
any embedding tensor that satisfies the above constraints will lead to
a consistent gauge invariant and supersymmetric theory, irrespective
of whether the charges are electric or magnetic.

There are two characteristic features that play an important role in
this paper.  One that is typical of four-dimensional theories with
vector gauge fields, concerns electric/magnetic duality
\cite{deWit:2001pz}.  For zero gauge-coupling constant, the gauge
fields transform in the ${\bf 56}$ representation of \Exc7, and
decompose into 28 electric gauge fields and their 28 magnetic duals.
The magnetic duals do not appear in the Lagrangian, so that the
Lagrangian cannot be invariant under \Exc7, but the combined equations
of motion and Bianchi identities of the vector fields do transform
covariantly in the ${\bf 56}$ representation \cite{cremmer}. In fact
the rigid symmetry group of the Lagrangian is a subgroup of \Exc7
under whose action electric gauge fields are transformed into electric
gauge fields. This group is not unique. It depends on the embedding of
\Exc7 inside the larger duality group ${\rm Sp}(56;\mathbb{R})$, which
determines which gauge fields belonging to the ${\bf 56}$
representation play the role of electric and which ones the role of
magnetic gauge fields. The choice of the electric/magnetic frame fixes
the rigid symmetry group of the ungauged Lagrangian, and different
choices yield in general different Lagrangians which are not related
to each other by local field redefinitions.  

The conventional approach for introducing local gauge invariance
relies on the rigid symmetry group of the ungauged Lagrangian as the
gauge group has to be a subgroup thereof. The procedure requires the
introduction of minimal couplings involving only the electric vector
fields, and therefore it explicitly breaks the original \Exc7 duality
covariance of the field equations and Bianchi identities. The
advantage of the formulation proposed in \cite{dWST6} is two-fold. On
the one hand, minimal couplings involve both electric and magnetic
vector fields in symplectically invariant combinations with the
corresponding components of the embedding tensor. This ensures that,
irrespective of the gauge group, the \Exc7 invariance can formally be
restored at the level of the field equations and Bianchi identities,
provided the embedding tensor is treated as a ``spurionic'' object
transforming under \Exc7 and subject to the two aforementioned
group-theoretical constraints.  On the other hand, we are no longer
restricted in the choice of the gauge group by the rigid symmetries of
the original ungauged Lagrangian.  Regardless of the electric/magnetic
frame, we may introduce any gauge group contained in \Exc7
corresponding to an embedding tensor that satisfies the two
\Exc7-covariant constraints.  If this gauge group is not a subgroup of
the rigid symmetry group of the ungauged Lagrangian, the embedding
tensor will typically lead to magnetic charges and magnetic gauge
fields together with the tensor fields. The latter will play a crucial
role in realizing the gauge invariance of the final Lagrangian. An
interesting feature of the resulting theory is that the scalar
potential is described by means of a universal formula which is
independent of the electric/magnetic duality frame. 

A second, more general, feature of maximal supergravity is that the
scalar fields parametrize a symmetric space, in this case the coset
space ${\rm E}_{7(7)}/{\rm SU}(8)$. The standard treatment of the
corresponding gauged nonlinear sigma models is based on a formulation
in which the group ${\rm SU}(8)$ is realized as a local invariance
which acts on the spinor fields and the scalars; the corresponding
connections are composite fields. A gauging is based on a group ${\rm
  G}_g\subset {\rm E}_{7(7)}$ whose connections are provided by (some
of the) elementary vector gauge fields of the supergravity theory. The
coupling constant associated with the gauge group will be denoted by
$g$. One can impose a gauge condition with respect to the local ${\rm
  SU}(8)$ invariance which amounts to fixing a coset representative
for the coset space. In that case the \Exc7-symmetries will act
nonlinearly on the fields and these nonlinearities make many
calculations intractable. Because it is much more convenient to work
with symmetries that are realized linearly, the best strategy is
therefore to postpone this gauge fixing until the end. This strategy
was already adopted in \cite{deWitNic}, but in this paper we find it
convenient to introduce a slightly different definition of the coset
representative.

Let us end this introduction by making some remarks on the physical
significance of the embedding tensor. As previously anticipated, the
low-energy dynamics of any superstring/M-theory compactification that
admits a four dimensional $N=8$ effective supergravity description,
has to be contained within the class of theories discussed in the
present paper. From the higher-dimensional perspective a gauging is in
general characterized by constant background quantities which may be
related to fluxes of higher-dimensional field strengths across cycles
of the compactification manifold (form--fluxes), or just associated
with the geometry of the internal manifold (geometric--fluxes), such
as the tensor defining a twist in the topology in an internal torus
\cite{ScherkSchwarz}.  In all known instances of gauged extended
supergravities arising from superstring/M--theory compactifications,
these background quantities enter the four-dimensional theory as
components of the embedding tensor.  Interestingly enough, in these
cases the quadratic constraint on the embedding tensor follows from
consistency of the higher-dimensional field equations and Bianchi
identities. For instance, in type-II compactifications in the presence
of form-fluxes the quadratic constraint expresses the tadpole
cancellation condition.  This condition, in the context of
compactifications which are effectively described by $N=8$
four-dimensional supergravity, poses severe restrictions on the fluxes
since there is no room in this framework for localized sources such as
orientifold planes. This is the case, for example, for the type-IIB
theory compactified on a six-torus in the presence of NS-NS and R-R
form-fluxes. The situation is clearly different for compactifications
yielding $N\le 4$ theories in four dimensions.

Having identified the background quantities in a generic flux
compactification with components of the embedding tensor, our
formulation of gauged maximal supergravity may provide a useful
setting for studying the duality relations between more general
superstring/M-theory vacua. Indeed the embedding tensor transforms
covariantly with respect to the full rigid symmetry group \Exc7 of the
four-dimensional theory, which is expected to encode the various
string dualities. For instance, the generic T-duality transformations
on the string moduli of the six-torus, within the same type-II theory, are implemented by the
$\mathrm{SO}(6,6;\mathbb{Z})$ subgroup of \Exc7.

This paper is organized as follows. In section \ref{sec:embed} the
embedding tensor is introduced together with an extensive discussion
of the constraints it should satisfy. It is demonstrated in a special
electric/magnetic frame how these constraints ensure the existence of
a Lagrangian that is invariant under the gauge group specified by the
embedding tensor. Furthermore it is explained how to incorporate both
electric and magnetic charges and corresponding gauge fields. In
section \ref{sec:T-tensor} the corresponding $T$-tensor is introduced.
As a result of the constraints on the embedding tensor the $T$-tensor
satisfies a number of identities which are important for the
supersymmetry of the Lagrangian. In section \ref{sec:Lagr+transf} the
Lagrangian and the supersymmetry transformations are derived. Salient
features are the universal expressions for the fermionic masslike
terms and the scalar potential, which are induced by the gauging, as
well as the role played by the magnetic gauge fields. Some
applications, including explicit examples of new gaugings, are
reviewed in section~5.

\section{The embedding tensor}
\setcounter{equation}{0}
\label{sec:embed}
We start by considering (abelian) vector fields $A_\m{}^M$
transforming in the ${\bf 56}$ representation of the \Exc7 duality
group with generators denoted by $(t_\a)_M{}^N$, so that $\d A_\m{}^M
= -\Lambda^\a (t_\a)_N{}^M\,A_\m{}^N$. These vector potentials can be
decomposed into 28 electric potentials $A_\mu{}^\Lambda$ and 28
magnetic potentials $A_{\mu\Lambda}$. In the conventional supergravity
Lagrangians only 28 electric vectors appear, but at this stage we base
ourselves on 56 gauge fields. In due course we will see how the
correct balance of physical degrees of freedom is nevertheless
realized. The gauge group must be a subgroup of \Exc7, so that its
generators $X_M$, which couple to the gauge fields $A_\mu{}^M$, are
decomposed in terms of the 133 independent \Exc7 generators $t_\a$,
{\it i.e.},
\begin{equation}
  \label{eq:X-theta-t}
  X_M = \Theta_M{}^\a\,t_\a\;,
\end{equation}
where $\a= 1,2,\ldots,133$ and $M=1,2,\ldots,56$.  The gauging is thus
encoded in a real {\it embedding tensor} $\Theta_{M}{}^{\alpha}$
belonging to the ${\bf 56}\times{\bf 133}$ representation of \Exc7.
The embedding tensor acts as a projector whose rank $r$ equals the
dimension of the gauge group. One expects that $r\leq28$, because the
ungauged Lagrangian should be based on 28 vector fields to describe
the physical degrees of freedom. As we shall see shortly, this bound
is indeed satisfied. The strategy of this paper is to treat the
embedding tensor as a spurionic object that transforms under the
duality group, so that the Lagrangian and transformation rules remain
formally invariant under \Exc7. The embedding tensor can then be
characterized group-theoretically. When freezing $\Theta_M{}^\a$ to a
constant, the \Exc7-invariance is broken.  An admissible embedding
tensor is subject to a linear and a quadratic constraint, which ensure
that one is dealing with a proper subgroup of \Exc7 and that the
corresponding supergravity action remains supersymmetric. These
constraints are derived in the first subsection. A second subsection
elucidates some of the results in a convenient \Exc7 basis. A third
subsection deals with the introduction of tensor gauge fields and
their relevance for magnetic charges.
\subsection{The constraints on the embedding tensor}
\label{sec:constraints}
The fact that the $X_M$ generate a group and thus define a Lie algebra,
\be
\label{gauge-algebra}
{[X_M,X_N]} = f_{MN}{}^P\,X_P,
\ee
with $f_{MN}{}^P$ the as yet unknown structure constants of the gauge group,
implies that the embedding tensor must satisfy the closure condition,
\begin{equation}
\label{eq:gauge-gen}
\Theta_M{}^\a\,\Theta_N{}^\b \,f_{\a\b}{}^{\g}= f_{MN}{}^P\,
\Theta_P{}^\g\,.
\end{equation} 
Here the $f_{\a\b}{}^\g$ denote the structure constants of \Exc7,
according to $[t_\a,t_\b]= f_{\a\b}{}^\g\,t_\g$. The
closure condition implies that the structure constants $f_{MN}{}^P$ satisfy the
Jacobi identities in the subspace projected by the embedding tensor,
\begin{equation}
\label{eq:jacobi}
f_{[MN}{}^Q\,f_{P]Q}{}^R\,\Theta_R{}^\a =0\,.
\end{equation}
Using the gauge group generators $X_M$ one introduces gauge covariant
derivatives,  
\begin{equation}
\label{eq:cov-der}
D_\m = \partial_\m - g\,A_\m{}^M\,X_M\,, 
\end{equation}
where $g$ denotes an uniform gauge coupling constant. These derivatives lead
to the covariant field strengths, 
\begin{equation}
  \label{eq:Theta-field-strength}
\Theta_M{}^\a \,{\cal F}_{\m\n}{}^M = \Theta_M{}^\a \,(\pa_\m A_\n{}^M  -
\pa_\n A_\m{}^M   -g\,f_{NP}{}^M \, A_\m{}^N A_\n{}^P)\,.    
\end{equation}
The gauge field transformations are given by 
\begin{equation}
  \label{eq:gauge-field-transf}
   \Theta_M{}^\a \,\d
   A_\m{}^M = \Theta_M{}^\a\, (\pa_\mu \Lambda^M - g\, f_{NP}{}^M
   \,A_\m{}^N\,\Lambda^P)\,.
\end{equation}
Because of the contraction with the embedding tensor, the above
results apply only to an $r$-dimensional subset of the gauge fields;
the remaining ones do not appear in the covariant derivatives and are
not directly involved in the gauging.  However, the $r$ gauge fields
that do appear in the covariant derivatives, are only determined up to
additive terms linear in the $56-r$ gauge fields that vanish upon
contraction with $\Theta_M{}^\a$.

While the gauge generators \eqn{eq:X-theta-t} act in principle uniformly on all
fields that transform under \Exc7, the gauge field transformations are a bit
more subtle to determine. This is so because the gauge fields involved in the
gauging should transform in the adjoint representation of the gauge
group. At the same time their charges should 
coincide with $X_M$ in the ${\bf 56}$ representation, so that $(X_M)_N{}^P$ 
must decompose into the adjoint representation of the gauge 
group plus possible extra terms which vanish upon contraction with the
embedding tensor, 
\begin{equation}
  \label{eq:adjoint}
 (X_M)_N{}^P \,\Theta_P{}^\a \equiv \Theta_M{}^\b \,t_{\b N}{}^P\,
 \Theta_P{}^\a = - f_{MN}{}^P \,\Theta_P{}^\a \,.  
\end{equation} 
These extra terms, pertaining to the gauge fields that do not appear
in the covariant derivatives, will be considered in due course. Note
that \eqn{eq:adjoint} is the analogue of (\ref{eq:gauge-gen}) in the
${\bf 56}$ representation.  The combined conditions \eqn{eq:gauge-gen}
and \eqn{eq:adjoint} imply that $\Theta$ is invariant under the gauge
group and yield the \Exc7-covariant condition
\begin{equation}
    \label{closure-constraint}
    C_{MN}{}^\a \equiv f_{\b\g}{}^\a \, \Theta_M{}^\b\,\Theta_N{}^\g +
    t_{\b N}{}^P\,
    \Theta_M{}^\b \, \Theta_P{}^\a =0 \;.  
\end{equation}
Obviously $C_{MN}{}^\alpha$ can be assigned to irreducible \Exc7
representations contained in the 
${\bf 56}\times{\bf 56}\times{\bf 133}$ representation. The condition
\eqn{closure-constraint} encompasses all previous results: it implies that 
\begin{equation}
{[X_M,X_N]} = -X_{MN}{}^P\,X_P,
\end{equation}
so that \eqn{closure-constraint} implies a closed gauge algebra, whose
structure constants, related to $X_{MN}{}^P$ in accord with
\eqn{eq:adjoint}, have the required antisymmetry. Hence
\eqn{closure-constraint}
is indeed sufficient for defining a proper subgroup embedding.\footnote{
 Note that for an abelian gauge group we have
 $X_{MN}{}^P\Theta_P{}^\a=0$. Using \eqn{eq:repres-constraint} this leads to
 ${\rm tr}(X_M\,X_N) =0$.}  

The embedding tensor satisfies a second constraint, which is required by
supersymmetry. This constraint is 
linear and amounts to restricting $\Theta_M{}^\a$ to the ${\bf 912}$
representation \cite{dWST1}. From  
\begin{equation}
  \label{eq:branching}
  {\bf 56}\times{\bf 133} =
  {\bf 56} + {\bf 912} + {\bf 6480}\,,
\end{equation}
one shows that this condition on the representation implies the equations,
\begin{equation}
    \label{eq:repres-constraint}
    t_{\alpha M}{}^N\,\Theta_N{}^\alpha = 0\,,\qquad
   (t_{\beta} t^{\alpha})_ M{}^N\,\Theta_N{}^\beta = -\ft12\,
    \Theta_M{}^\alpha \,,
\end{equation}
where the index $\a$ is raised by the inverse of the \Exc7-invariant
metric $\eta_{\alpha\beta}= {\rm tr}(t_\a t_\b)$. 

As a result of the representation constraint, the
representation content of $C_{MN}{}^\alpha$ can be further restricted as from
\eqn{eq:repres-constraint} one can derive the following equations,
\begin{equation}
  \label{eq:C-constraints}
t_{\alpha N}{}^P\, C_{MP}{}^\alpha =0 \,, \quad
(t_\beta \,t^{\alpha})_ N{}^P\, C_{MP}{}^\beta =
- \ft12\,C_{MN}{}^\alpha \,,\quad
t_{\alpha M}{}^P\, C_{PN}{}^\alpha= t_{\alpha N}{}^P\,
C_{PM}{}^\alpha\,.
\end{equation}
They imply that $C_{MN}{}^\alpha$ should belong to representations
contained in ${\bf 56}\times{\bf 912}$. On the other hand, the product of two
$\Theta$-tensors belongs to the symmetric product of two ${\bf 912}$
representations. Comparing the decomposition of these two
products\footnote{
  We used the LiE package \cite{LeCoLi92} for computing the decompositions of
  tensor products and the branching of representations.}, 
\begin{eqnarray}
({\bf 912}\times {\bf 912})_{\rm s}&=&{\bf 133}+{\bf 8645}+
{\bf   1463}+{\bf 152152}+ {\bf 253935}\,, \nonumber\\ 
{\bf 56}\times {\bf 912} &=&{\bf 133}+{\bf 8645}+ {\bf 1539} + 
{\bf 40755}\,, 
\end{eqnarray}
one deduces that $C_{MN}{}^\a$ belongs to the ${\bf 133}+{\bf 8645}$
representation. Noting the decomposition $({\bf 133}\times{\bf 133})_{\rm a} 
={\bf 133}+{\bf 8645}$, we observe that there is an alternative way to
construct these two representations which makes use of the fact that 
${\rm Sp}(56;\mathbb{R})$, and thus its \Exc7 subgroup, has
an invariant skew-symmetric matrix $\Omega^{MN}$, which we write as, 
\begin{equation}
\label{eq:symplectic-omega}
\Omega^{MN} = \pmatrix{0&\bf{1}\cr -{\bf 1}& 0\cr}\;.  
\end{equation} 
The conjugate matrix $\Omega_{MN}$ takes the same form, so that
$\Omega^{MN}\Omega_{NP}= - \delta^M{}_P$. In this way one derives an
equivalent version of the constraint \eqn{closure-constraint},
\begin{equation}
  \label{eq:closure-constraint2}
  \Theta_M{}^\a \Theta_N{}^\b \,\Omega^{MN} = 0 
  \;\;\Longleftrightarrow\; \; 
\Theta^{\Lambda\,[\alpha}\Theta_{\Lambda}{}^{\beta]} =0 \;, 
\end{equation}
which is only equivalent provided the representation constraint
\eqn{eq:repres-constraint} is imposed. The constraint
\eqn{eq:closure-constraint2} implies that the $\Theta^\alpha$ can all
be chosen as ${\it electric}$ vectors upon a suitable ${\rm
  Sp}(56;\mathbb{R})$ transformation, implying that all the nonzero
components of the 133 vectors $\Theta^\a$ cover an $r$-dimensional
subspace parametrized by the gauge fields $A_\m{}^M$ with
$M=1,\ldots,r$ and $r\leq 28$. In this basis $X_{MN}{}^P$ can be
written in triangular form, 
\begin{equation}
\label{blockmatrix}
X_M= \pmatrix{ - f_M & a_M\cr \noalign{\vskip 1mm}0& b_M\cr}\,, 
\end{equation}
where the $r\times r$ upper-left diagonal block coincides with the
gauge group structure constants and the submatrices $a_M$ and $b_M$ do
not contribute to the product $(X_M)_N{}^P\Theta_P{}^\a$. The
lower-left $(56-r)\times r$ block vanishes as a result of
\eqn{eq:adjoint}.  It is easy to see that $a_M$ and $b_M$ cannot both
be zero. If that were the case, we would have $f_{MN}{}^P = -
X_{MN}{}^P$, which is antisymmetric in $M$ and $N$. Hence,
\begin{equation}
  \label{eq:antisymmetry}
  \Theta_N{}^\alpha\,t_{\alpha\,M}{}^P=
  - \Theta_M{}^\alpha\,t_{\alpha\,N}{}^P\,.
\end{equation}
Contracting this result by $(t^\beta)_P{}^M$ leads to
$t_\alpha\,t^\beta \, \Theta^\alpha = - \Theta^\beta$ which is in
contradiction with the representation constraint
\eqn{eq:repres-constraint}. In the next subsection we give a more
detailed analysis of the submatrices $a_M$ and $b_M$, which shows that
$b_M$ never vanishes.

Let us now proceed and find the restrictions on $X_{MN}{}^P$. First of
all, \Exc7 invariance of $\Omega^{MN}$ implies that $X_{MNP}=
X_{MN}{}^Q\Omega_{PQ}$ is symmetric in $N$ and $P$. Furthermore, $X$
belongs to the ${\bf 912}$ representation (remember that
$(t_\a)_M{}^N$ transforms as an \Exc7 invariant tensor, so that $X_M$
transforms in the same representation as the embedding tensor), which
is, however, {\it not} contained in the symmetric product $({\bf
  56}\times {\bf 56}\times {\bf 56})_{\rm s}$.  Consequently it
follows that the fully symmetric part of $X_{MNP}$ must vanish. Likewise,
contractions of $X_{MN}{}^P$ will also vanish, as they do not
correspond to the ${\bf 912}$ representation. Hence $X_{MN}{}^P$
has the following properties,
\begin{equation}
  \label{eq:X-condition}
  X_{M[NP]}= 0\,,\qquad X_{(MNP)} =0\,,\qquad X_{MN}{}^N =
  X_{MN}{}^M=0\,.    
\end{equation}
The first condition implies that 
\begin{equation}
  \label{eq:sympl-X}
  X_{M\Lambda}{}^\Sigma= -X_{M}{}^\Lambda{}_\Sigma\,,\quad  
  X_{M \Lambda\Sigma}=X_{M\Sigma\Lambda}\,, \quad
  X_{M}{}^{\Lambda\Sigma}=X_{M}{}^{\Sigma\Lambda}\,,   
\end{equation}
whereas the second one implies 
\begin{equation}
  \label{eq:lin}
  \begin{array} {l}
  X^{(\Lambda\Sigma\Gamma)}=0\,,\\
    X_{(\Lambda\Sigma\Gamma)}=0\,,
  \end{array}
  \qquad
  \begin{array} {l}
  2X^{(\Gamma\Lambda)}{}_{\Sigma}= 
  X_{\Sigma}{}^{\Lambda\Gamma}\,,\\
  2X_{(\Gamma\Lambda)}{}^{\Sigma}= X^{\Sigma}{}_{\Lambda\Gamma}\,.
  \end{array}
\end{equation}
The constraints (\ref{eq:sympl-X}) and (\ref{eq:lin}) coincide with
the constraints that we have adopted in a more general
four-dimensional context in \cite{dWST6}.

The constraint (\ref{eq:closure-constraint2}) motivates the
definition of another tensor $Z^{M,\alpha}$, which is orthogonal to
the embedding tensor, {\it i.e.} $Z^{M,\alpha}\Theta_M{}^\beta=0$, 
\begin{equation}
  \label{eq:def-Z}
  Z^{M,\alpha}\equiv \ft12 \Omega^{MN}\,\Theta_N{}^\a \quad
\Longrightarrow
\quad
\left\{
\begin{array}{rcr}
Z^{\Lambda\alpha} &=& \ft12\Theta^{\Lambda\alpha} \,,\\[1ex]
Z_{\Lambda}{}^{\alpha} &=& -\ft12\Theta_{\Lambda}{}^{\alpha} \,.
\end{array}
\right.
\end{equation}
As a consequence of the second equation of (\ref{eq:X-condition}), one
may derive, 
\begin{equation}
  \label{eq:Z-d}
  X_{(MN)}{}^{P}= Z^{P,\alpha}\, d_{\alpha \,MN} \;, 
\end{equation}
where $d_{\alpha \,MN}$ is an \Exc7-invariant tensor symmetric
in $(MN)$, 
\begin{equation}
  \label{eq:def-d}
d_{\alpha\, MN} \equiv (t_\alpha)_M{}^P\, \Omega_{NP}\,.
\end{equation}
The more general significance of (\ref{eq:Z-d}) was discussed in
\cite{deWit:2005hv}. 

\subsection{A special \Exc7 basis}
\label{sec:special-basis}
To appreciate the various implications of the constraints on
$X_{MN}{}^P$, we consider a special basis in which all the charges are
electric. Hence magnetic charges vanish by virtue of
$\Theta^{\Lambda\alpha}=0$. A vector $V^M$ in the ${\bf 56}$
representation can then be decomposed according to
\begin{equation}
  \label{eq:special-basis}
  V^M \longrightarrow (V^\Lambda,V_\Lambda) \longrightarrow (V^A, V^a,
  V_{A}, V_{a})\;,
\end{equation}
with $A= 1,\ldots,r$ and $a=r+1,\ldots, 28$; {\it i.e.}, electric
(gauge field) components are written with upper indices $A,a$ and
their magnetic duals with corresponding lower indices $A,a$. The
components $V^a$ then span the subspace defined by the condition
$\Theta_\Lambda{}^\a\,V^\Lambda=0$. Consequently, $V^A$ and $V_A$ are
defined up to terms proportional to the $V^a$ and $V_a$, respectively.
Obviously only the $\Theta_A{}^\alpha$ are nonvanishing and the
$X_{MN}{}^P$ are only nonzero when $M=A$. Imposing (\ref{eq:Z-d}) and
(\ref{eq:def-d}), it follows that a block decomposition of
$X_{AN}{}^P$ is then as follows (row and column indices are denoted by
$B,b$ and $C,c$, respectively),
\begin{equation}
  \label{eq:X-decomposition}
  X_{AN}{}^P = 
  \pmatrix{ -f_{AB}{}^C & h_{AB}{}^c &  C_{ABC}    & C_{ABc}\cr
    \noalign{\vskip 1.5mm}
                  0   & 0          &  C_{ACb}    & 0      \cr
    \noalign{\vskip 1.5mm}
                  0   & 0          &  f_{AC}{}^B & 0      \cr
    \noalign{\vskip 1.5mm}
                  0   & 0          & -h_{AC}{}^b & 0      \cr} \;,
\end{equation}
where 
\be
h_{(AB)}{}^c =C_{(AB)c}= C_{A[BC]}= C_{(ABC)} = f_{(AB)}{}^C= f_{AB}{}^B=0\,.
\ee
The last equation implies that the gauge group is unimodular.
The closure relations \eqn{gauge-algebra} imply a number of
nontrivial identities, 
\begin{eqnarray}
\label{eq:group-ids}
f_{[AB}{}^D\,f_{C]D}{}^E &=&0\,,\nonumber \\
f_{[AB}{}^D\,h_{C]D}{}^a &=&0\,,\nonumber \\
f_{AB}{}^E\,C_{ECD} - 4\, f_{(C[A}{}^E\,C_{B]D)E}
+ 4\, h_{(C[A}{}^a\,C_{B]D)a}&=&0\,,\nonumber\\
f_{[AB}{}^D\,C_{C]D a} &=&0\,.  
\end{eqnarray}
The transformations generated by (\ref{eq:X-decomposition}) imply that
electric gauge fields transform exclusively into electric gauge fields, 
\begin{eqnarray}
  \label{eq:delta-A-electric}
  \delta A_\mu{}^A &=& \Lambda^B\, f_{BC}{}^A  \,A_\mu{}^C \,,
  \nonumber \\
  \delta A_\m{}^a &=&  - \Lambda^B\,h_{BC}{}^a \,A_\mu{}^C \,,
\end{eqnarray}
where, for the moment, we keep the transformation parameters
$\Lambda^A$ space-time independent. The magnetic gauge fields, on the
other hand, transform into electric and magnetic gauge fields, 
\begin{eqnarray}
  \label{eq:delta-A-magnetic}
  \delta A_{\mu A} &=&- \Lambda^B( f_{BA}{}^C \,A_{\mu\, C} -
  h_{BA}{}^c \, A_{\mu\,c} +  C_{BAC} \,A_{\mu}{}^C + C_{BAc}\,
  A_{\mu}{}^c) \,,   \nonumber \\
  \delta A_{\mu\,a} &=&{} - \Lambda^B\, C_{BAa}  \,A_{\mu}{}^A \,. 
\end{eqnarray}
Because the Lagrangian does not contain the magnetic gauge fields in
this case, the question arises how the gauge transformations are
realized.  The answer is provided by electric/magnetic duality. The
above variations (\ref{eq:delta-A-electric}) and
(\ref{eq:delta-A-magnetic}) generate a subgroup of these duality
transformations that must be contained in \Exc7. General
electric/magnetic transformations constitute an even bigger group
${\rm Sp}(56;\mathbb{R})$. In the abelian case they are defined by
rotations of the 28 field strengths $F_{\mu\nu}{}^\Lambda$ and the 28
conjugate tensors ${G}_{\mu\nu\,\Lambda}$ defined by
\begin{equation}
  \label{eq:def-G}
  {G}_{\mu\nu\,\Lambda} = \mathrm{i} \, 
  \varepsilon_{\mu\nu\rho\sigma}\,
  \frac{\partial \mathcal{L}}{\partial{F}_{\rho\sigma}{}^\Lambda}
  \;.
\end{equation}
The corresponding field equations and Bianchi identities constitute 56
equations,
\begin{equation}
  \label{eq:eom-bianchi}
  \partial_{[\mu} {F}_{\nu\rho]}{}^\Lambda  
   = 0 = \partial_{[\mu}{G}_{\nu\rho]\,\Lambda} \,,
\end{equation}
which are clearly invariant under rotations of the 56 field strengths 
$G_{\mu\nu}{}^M$, defined by
\begin{equation}
  \label{eq:G-M}
  G_{\mu\nu}{}^M\equiv \pmatrix{F_{\mu\nu}{}^\Lambda \cr
    \noalign{\vskip 1mm}
     G_{\mu\nu \Sigma}}\;. 
\end{equation}
The equations (\ref{eq:eom-bianchi}) show that the $G_{\mu\nu}{}^M$
can be expressed in terms of 56 vector potentials, and this is how the
electric and magnetic gauge fields appear in the abelian case. Hence
we may write,
\begin{equation}
  \label{eq:vector-potentials}
  G_{\mu\nu}{}^M = 2 \,\partial_{[\mu}A_{\nu]}{}^M\,.
\end{equation}
Electric/magnetic duality acts in principle on (abelian) field
strengths rather than on corresponding gauge fields, because the field
strengths $G_{\mu\nu\Lambda}$ are not independent according to
(\ref{eq:def-G}).

Let us briefly return to the general ${\rm Sp}(56;\mathbb{R})$ dualities,
which can be decomposed as follows,
\begin{equation}
  \label{eq:em-duality}
  \pmatrix{{F}^\Lambda\cr   \noalign{\vskip 1.5mm} {G}_\Lambda}
  \longrightarrow  
  \pmatrix{U^\Lambda{}_\Sigma & Z^{\Lambda\Sigma} \cr
    \noalign{\vskip 1.5mm}
    W_{\Lambda\Sigma} & V_\Lambda{}^\Sigma }   
  \pmatrix{{F}^\Sigma\cr \noalign{\vskip 1.5mm}  {G}_\Sigma} \,,
\end{equation}
where the (real) constant matrix leaves the skew-symmetric matrix
$\Omega_{MN}$ invariant. This ensures that the new dual field
strengths ${G}_{\mu\nu\Lambda}$ can again be written in the form
(\ref{eq:def-G}) but with a different Lagrangian. These duality
transformations thus define equivalence classes of Lagrangians that
lead to the same field equations and Bianchi identities. They are
generalizations of the duality transformations known from Maxwell
theory, which rotate the electric and magnetic fields and inductions
(for a review of electric/magnetic duality, see \cite{deWit:2001pz}).
An \Exc7 subgroup of these transformations, combined with
transformations on the scalar fields, constitutes an {\em invariance}
group, meaning that the combined field equations and Bianchi
identities (including the field equations for the other fields) before
and after the \Exc7 transformation follow from an identical
Lagrangian. Only the vector field strengths (\ref{eq:G-M}) and the
scalar fields (to be introduced in section \ref{sec:T-tensor}) are
subject to these \Exc7 transformations.  The other fields, such as the
vierbein field and the spinor fields, are inert under \Exc7.

To be more specific let us introduce the generic gauge field Lagrangian
that is at most quadratic in the field strengths, parametrized as in
\cite{dWST6},
\begin{eqnarray}
  \label{eq:quadratic-L}
  e^{-1} \mathcal{L}_{\rm vector} &=&
-\ft14 \, \mathrm{i} \Big\{\mathcal{N}_{\Lambda\Sigma} 
\, {F}_{\mu\nu}^+{}^{\Lambda}\, {F}^{+\mu\nu\Sigma}  -
 \bar\mathcal{N}_{\Lambda\Sigma} \, {F}_{\mu\nu}^-{}^{\Lambda}\, 
  {F}^{-\mu\nu\Sigma} \Big\} \nonumber \\[1ex]
&&{} 
  + {F}_{\mu\nu}^{+\Lambda}
  \,\mathcal{O}_\Lambda^{+\mu\nu} + {F}_{\mu\nu}^{-\Lambda}
  \,\mathcal{O}_\Lambda^{-\mu\nu}  \nonumber\\[1ex]
  &&{} 
  + \mathrm{i} [(\mathcal{N} -\bar\mathcal{N})^{-1}]^{\Lambda\Sigma} \;\Big[
  \mathcal{O}_{\mu\nu\Lambda}^+ \mathcal{O}^{+\mu\nu}_{\Sigma}+ 
  \mathcal{O}_{\mu\nu\Lambda}^- \mathcal{O}^{-\mu\nu}_{\Sigma}\Big]\,.
\end{eqnarray}
Here the ${F}^\pm_{\mu\nu}$ are complex (anti-)selfdual combinations
normalized such that ${F}_{\mu\nu}= {F}^+_{\mu\nu} + {F}^-_{\mu\nu}$.
The field-dependent symmetric tensor $\mathcal{N}_{\Lambda\Sigma}$
comprises the generalized theta angles and coupling constants and
$\mathcal{O}_{\mu\nu\,\Lambda}^\pm$ represents bilinears in the
fermion fields. The terms quadratic in
$\mathcal{O}_{\mu\nu}{}^\Lambda$ are such that any additional terms in
the Lagrangian (which no longer depends on the field strengths)
will transform covariantly under electric/magnetic duality. From the
above Lagrangian we derive
\begin{equation}
  \label{eq:G-F-relation}
  G^+_{\mu\nu\Lambda} = \mathcal{N}_{\Lambda\Sigma}\, 
  F^+_{\mu\nu}{}^\Sigma + 2\mathrm{i} \,\mathcal{O}^+_{\mu\nu \Lambda} \,.
\end{equation}
Upon an electric/magnetic duality transformation (\ref{eq:em-duality})
one finds an alternative Lagrangian of the same form but with a
different expression for $\mathcal{N}_{\Lambda\Sigma}$ and
$\mathcal{O}_\Lambda$,
\begin{eqnarray}
  \label{eq:N-O-transform}
  \mathcal{N}_{\Lambda\Sigma}&\longrightarrow&
  (V \mathcal{N} + W )_{\Lambda\Gamma} \,
  [(U + Z \mathcal{N})^{-1}]^\Gamma{}_\Sigma \,,\nonumber\\
   \mathcal{O}^{+}_{\mu\nu\,\Lambda} &\longrightarrow&
   \mathcal{O}^{+}_{\mu\nu\,\Sigma} \, [(U +
   Z\mathcal{N})^{-1}]^\Sigma{}_\Lambda\,, 
\end{eqnarray}
This result follows from requiring consistency between
(\ref{eq:def-G}) and (\ref{eq:em-duality}). The restriction to ${\rm
  Sp}(56;\mathbb{R})$ ensures that the symmetry of
$\mathcal{N}_{\Lambda\Sigma}$ remains preserved. For the \Exc7
subgroup of invariances, the transformations (\ref{eq:N-O-transform})
must be induced by corresponding \Exc7 transformations of the scalar
fields.

Let us now return to the infinitesimal gauge transformations
corresponding to the charges (\ref{eq:X-decomposition}), which act on
the field strengths according to $\delta F_{\mu\nu}{}^M = - \Lambda^A
\,X_{AN}{}^M\, F_{\mu\nu}{}^N$. The abelian field strengths
$F_{\mu\nu}{}^\Lambda$ and $G_{\mu\nu\Lambda}$ thus transform as
\begin{eqnarray}
  \label{eq:delta-F-G}
  \d F_{\mu\nu}{}^A &=& \Lambda^B\, f_{BC}{}^A  \,F_{\mu\nu}{}^C \,,
  \nonumber \\
  \d F_{\mu\nu}{}^a &=&  - \Lambda^B\,h_{BC}{}^a \,F_{\mu\nu}{}^C \,,
  \nonumber \\
  \delta G_{\mu\nu\,A} &=&- \Lambda^B( f_{BA}{}^C \,G_{\mu\nu\, C} -
  h_{BA}{}^c \, 
  G_{\m\n\,c} +  C_{BAC} \,F_{\m\n}{}^C + C_{BAc}\, F_{\mu\nu}{}^c) \,,
  \nonumber \\
  \delta G_{\mu\nu\,a} &=& - \Lambda^B\, C_{BAa}  \,F_{\mu\nu}{}^A \,. 
\end{eqnarray}
According to (\ref{eq:G-F-relation}) the field strengths
$G_{\mu\nu\Lambda}$ depend also on fields other than the vector
fields, and in order to have an invariance, the transformations of the
these fields should combine with the transformations of the vector
fields to yield the above variations for the dual field strengths
$G_{\mu\nu\Lambda}$. Therefore the gauge group must be a subgroup of
\Exc7. In that case it follows that the transformations
(\ref{eq:delta-F-G}) for $F_{\mu\nu}{}^\Lambda$ and
$G_{\mu\nu\Lambda}$ leave the Lagrangian (\ref{eq:quadratic-L})
invariant, up to
\begin{equation}
\label{abelian-variation}
\d{\cal L}\propto     \varepsilon^{\m\n\rho\sigma}\,  \Lambda^A\,
\Big[ C_{ABC}\, F_{\m\n}{}^B\,F_{\rho\sigma}{}^C + 2\, C_{ABa}\,
F_{\m\n}{}^B\,F_{\rho\sigma}{}^a \Big] \,.
\end{equation}
This variation constitutes a total derivative when the $\Lambda^A$ are
constant.  When the parameters $\Lambda^A$ are space-time dependent,
one needs to introduce extra terms into the Lagrangian. According to
\eqn{eq:X-decomposition} the gauge fields transform as, 
\begin{eqnarray}
  \label{eq:gauge-fields-trans}
  \delta  A_\m{}^A &=& \pa_\mu \Lambda^A - g\,
  f_{BC}{}^A \,A_\m{}^B\,\Lambda^C \,, \nonumber\\
  \delta A_\m{}^a &=& \pa_\mu \Lambda^a + g\,
  h_{BC}{}^a \,A_\m{}^B\,\Lambda^C \,, 
\end{eqnarray}
and the covariant field strengths acquire the standard non-abelian
modifications,  
\begin{eqnarray}
\label{eq:cov-field-strengths}
F_{\mu\nu}{}^A\to {\cal F}_{\m\n}{}^A &=& \pa_\m A_\n{}^A  - \pa_\n
A_\m{}^A   -g\,f_{BC}{}^A \, A_\m{}^B A_\n{}^C \,,  \nonumber\\
F_{\mu\nu}{}^a \to {\cal F}_{\m\n}{}^a &=& \pa_\m A_\n{}^a  - \pa_\n
A_\m{}^a   + g\,h_{BC}{}^a \, A_\m{}^B A_\n{}^C \,.  
\end{eqnarray} 
Likewise the derivatives on the scalar fields are extended to properly  
covariantized derivatives according to \eqn{eq:cov-der}. 
The only gauge fields that appear in the covariant derivatives
are the fields $A_\m{}^A$, so that only these gauge fields couple to the matter
fields. Note that, according to \eqn{eq:gauge-fields-trans} and
\eqn{eq:cov-field-strengths}, the abelian gauge fields $A_\m{}^a$ couple to
charges that are central in the gauge algebra. Therefore the resulting gauge
algebra is a central extension of \eqn{gauge-algebra}. Introducing formal
generators $\tilde X_A$ and $\tilde X_a$, it reads,
\begin{equation}
{[}\tilde X_A,\tilde X_B]= f_{AB}{}^C\,\tilde X_C - h_{AB}{}^a \tilde X_a\,. 
\end{equation} 
On the matter fields the central charges $\tilde X_a$ vanish and $\tilde X_A=
X_A$. 

In \eqn{abelian-variation} the abelian field
strengths will be replaced by the covariant field strenths
\eqn{eq:cov-field-strengths}, so that \eqn{abelian-variation} is no longer a
total derivative. Therefore the invariance of the action requires the presence
of extra Chern-Simons-like terms, 
\begin{eqnarray}
  \label{eq:Chern-Simons}
  {\cal L}^{\rm CS}&\propto&  g\, \varepsilon^{\m\n\rho\sigma}\,  
  \Big[ C_{ABC}\, A_\m{}^A
  A_\n{}^B (\pa_\rho A_\sigma{}^C - \ft38 g\,f_{DE}{}^C A_\rho{}^D
  A_\sigma{}^E)  \nonumber\\  
  &&{}\hspace{14mm}
  +  C_{ABa} \,A_\m{}^A  (A_\n{}^B \pa_\rho A_\sigma{}^a + A_\n{}^a\pa_\rho
  A_\sigma{}^B) \nonumber\\  
  &&{}\hspace{14mm}
  +\ft38 g\, C_{ABa} \,A_\m{}^A  ( h_{CD}{}^a A_\n{}^B  - f_{CD}{}^B \,
  A_\n{}^a)  A_\rho{}^C  A_\sigma{}^D \Big] \;.
\end{eqnarray}  
The identities \eqn{eq:group-ids} ensure that these terms are indeed sufficient
for restoring the gauge invariance of the Lagrangian
\cite{dWLVP,dWHR}. In this connection it is important that the
definition of the dual field strengths remains as in (\ref{eq:def-G}),
so that $\mathcal{G}_{\mu\nu\Lambda}$ will be defined by
(\ref{eq:G-F-relation}) with $F_{\mu\nu}{}^\Lambda$ replaced by the
non-abelian field strengths $\mathcal{F}_{\mu\nu}{}^\Lambda$ defined in
(\ref{eq:cov-field-strengths}). 

Hence we have shown that any embedding tensor that satisfies the two
constraints \eqn{closure-constraint} and \eqn{eq:repres-constraint},
leads to a gauge invariant Lagrangian. We emphasize once more that
this was done in the special basis \eqn{eq:special-basis}, in which
the charges are electric. The magnetic gauge fields do not play a role
here and in the non-abelian case they can no longer be defined in
terms of a solution of (\ref{eq:vector-potentials}).

\subsection{Magnetic potentials and antisymmetric tensor fields}
\label{sec:antisymmetric-tensors} 
In the more general setting with magnetic charges, the gauge algebra
does not close, simply because the Jacobi identity is only valid on
the subspace projected by the embedding tensor ({\it c.f.}
(\ref{eq:jacobi})). As was generally proven in \cite{dWST6} for
four-dimensional gauge theories, one can still obtain a consistent
gauge algebra, provided one introduces magnetic gauge fields from the
beginning, together with tensor gauge fields $B_{\mu\nu\alpha}$. In
the case at hand these fields transform in the adjoint ${\bf 133}$
representation of \Exc7. At the same time, to avoid
unwanted degrees of freedom, the gauge transformations associated with
the tensor fields should act on the (electric and magnetic) gauge
fields by means of a transformation that also depends on the embedding
tensor,
\begin{equation}
  \label{eq:A-var}
\delta A_\mu{}^M =  D_\mu\Lambda^M- g\,Z^{M,\alpha}\,\Xi_{\m\,\alpha}\,,  
\end{equation}
where the $\Lambda^M$ are the gauge transformation parameters and the
covariant derivative reads, $D_\mu\Lambda^M =\partial_\mu\Lambda^M +
g\, X_{PQ}{}^M\,A_\mu{}^P\Lambda^Q$. The transformations proportional
to $\Xi_{\m\,\alpha}$ enable one to gauge away those vector fields
that are in the sector of the gauge generators $X_{MN}{}^P$ where the
Jacobi identity is not satisfied (this sector is perpendicular to the
embedding tensor). These gauge transformations form a group, as
follows from the commutation relations,
\begin{eqnarray}
  \label{eq:gauge-commutators}
  {}[\delta(\Lambda_1),\delta(\Lambda_2)] &=& \delta(\Lambda_3) +
  \delta(\Xi_3) \,,  \nonumber \\
  {}[\delta(\Lambda),\delta(\Xi)] &=& \delta(\tilde\Xi) \,, 
\end{eqnarray}
where 
\begin{eqnarray}
  \label{eq:gauge-parameters}
  \Lambda_3{}^M &=& g\,X_{[NP]}{}^M \Lambda_2^N\Lambda_1^P\,, \nonumber\\
  \Xi_{3 \mu\alpha} &=& d_{\alpha NP}( \Lambda_1^N
  D_\mu\Lambda_2^P - \Lambda_2^N D_\mu \Lambda_1^P) \,, \nonumber\\
  \tilde\Xi_{\mu\alpha} &=& g\Lambda^P( X_{P\alpha}{}^\beta + 2 d_{\alpha
  PN} Z^{N,\beta}) \Xi_{\mu\beta} \,. 
\end{eqnarray}

In order to write down invariant kinetic terms for the gauge fields we
have to define a suitable covariant field strength tensor. This is an
issue because the Jacobi identity is not satisfied and because we have
to deal with the new gauge transformations parametrized by the
parameters $\Xi_{\mu\alpha}$. Indeed, the usual field strength, which
follows from the Ricci identity, $[D_\mu,D_\nu]= - g
\mathcal{F}_{\mu\nu}{}^M\,X_M$,
\begin{equation}
  \label{eq:field-strength}
  {\cal  F}_{\mu\nu}{}^M =\pa_\m A_\n{}^M -\pa_\n A_\m{}^M + g\,
X_{[NP]}{}^M \,A_\m{}^N A_\n{}^P \,,
\end{equation}
is not fully covariant.\footnote{
   Observe that the covariant derivative is invariant under the tensor
   gauge transformations, so that the field strengths contracted with
   $X_M$ are in fact covariant. } 
The lack of covariance can be readily checked by observing that
$\mathcal{F}_{\mu\nu}{}^M$ does not satisfy the Palatini identity, 
\begin{equation}
  \label{eq:Palatini}
  \delta\mathcal{F}_{\mu\nu}{}^M = 2\, D_{[\mu}\delta A_{\nu]}{}^M -
  2 g\, X_{(PQ)}{}^M \,A_{[\mu}{}^P \,\delta A_{\nu]}{}^Q\,,
\end{equation}
under arbitrary variations $\delta A_\mu{}^M$. This result shows that
$\mathcal{F}_{\mu\nu}{}^M$ transforms under gauge transformations as
\begin{equation}
  \label{eq:delta-cal-F}
  \delta\mathcal{F}_{\mu\nu}{}^M= g\, \Lambda^P X_{NP}{}^M
  \,\mathcal{F}_{\mu\nu}{}^N - 2 g\, Z^{M,\alpha} (D_{[\mu}
  \Xi_{\nu]\alpha} +d_{\alpha PQ} \,A_{[\mu}{}^P\,\delta A_{\nu]}{}^Q)
  \,, 
\end{equation} 
which is not covariant. The standard strategy
\cite{dWST5,deWit:2005hv,dWST6} is therefore to define modified field
strengths,
\begin{equation}
  \label{eq:modified-fs}
{\cal H}_{\m\n}{}^M= 
{\cal F}_{\mu\nu}{}^M  + g\, Z^{M,\alpha} \,B_{\m\n\,\alpha}\;,
\end{equation}
where we introduce the tensor fields $B_{\m\n\,\alpha}$, which are
subject to suitably chosen gauge transformation rules. 

At this point we recall that the invariance transformations in the
rigid case implied that the field strengths $G_{\mu\nu}{}^M$ transform
under a subgroup of $\mathrm{Sp}(56,\mathbb{R})$ ({\it c.f.}
(\ref{eq:em-duality})). Our aim is to find a similar symplectic array
of field strengths so that these transformations are generated in the
non-abelian case as well. This is not possible based on the variations
of the vector fields $A_\mu{}^M$, which will never generate the type
of fermionic terms contained in $G_{\mu\nu\Lambda}$. However, the
presence of the tensor fields enables one to achieve this objective, at
least to some extent. Just as in the abelian case, we define an
$\mathrm{Sp}(56,\mathbb{R})$ array of field strengths
$\mathcal{G}_{\mu\nu}{}^M$ by
\begin{equation}
  \label{eq:cal-G-M}
  \mathcal{G}_{\mu\nu}{}^M\equiv \pmatrix{\mathcal{H}_{\mu\nu}{}^\Lambda \cr
    \noalign{\vskip 1mm}
     \mathcal{G}_{\mu\nu \Sigma}}\;,
\end{equation}
so that 
\begin{eqnarray}
  \label{eq:cal-G}
  \mathcal{G}^+_{\mu\nu}{}^\Lambda &=& \mathcal{H}^+_{\mu\nu}{}^\Lambda\,,
  \nonumber\\ 
  \mathcal{G}^+_{\mu\nu\Lambda} &=& \mathcal{N}_{\Lambda\Sigma}\, 
  \mathcal{H}^+_{\mu\nu}{}^\Sigma + 2\mathrm{i}
  \,\mathcal{O}^+_{\mu\nu \Lambda} \,. 
\end{eqnarray}
Note that the expression for $\mathcal{G}_{\mu\nu\Lambda}$ is the
analogue of (\ref{eq:G-F-relation}), with $F_{\mu\nu}{}^\Lambda$
replaced by $\mathcal{H}_{\mu\nu}{}^\Lambda$. 

Following \cite{dWST6} we introduce the following transformation rule for
$B_{\mu\nu\alpha}$ (contracted with $Z^{M,\alpha}$, because
only these combinations will appear in the Lagrangian),
\begin{equation}
  \label{eq:B-transf-0}
  Z^{M,\alpha} \delta B_{\mu\nu\,\alpha} = 2\,Z^{M,\alpha} 
   (D_{[\mu} \Xi_{\nu]\alpha} + d_{\alpha\,NP} A_{[\mu}{}^N \delta
   A_{\nu]}{}^P)  - 2\,X_{(NP)}{}^M  \Lambda^P
   \mathcal{G}_{\mu\nu}{}^N    \,,
\end{equation}
where $D_\mu \Xi_{\nu\alpha}= \partial_\mu \Xi_{\nu\alpha} - g
A_\mu{}^M X_{M\alpha}{}^\beta \Xi_{\nu\beta}$ with
$X_{M\alpha}{}^\beta= -\Theta_M{}^\gamma f_{\gamma\alpha}{}^\beta$ the
gauge group generators embedded in the adjoint representation of
\Exc7.  With this variation the modified field strengths
(\ref{eq:modified-fs}) are invariant under tensor gauge
transformations. Under the vector gauge transformations we derive the
following result,
\begin{eqnarray}
  \label{eq:delta-G/H}
  \delta \mathcal{G}^+_{\mu\nu}{}^\Lambda&=& - g\,\Lambda^P
  X_{PN}{}^\Lambda \,\mathcal{G}^+_{\mu\nu}{}^N  - g\,\Lambda^P
  X^\Gamma{}_P{}^\Lambda \,
  (\mathcal{G}^+_{\mu\nu} - \mathcal{H}^+_{\mu\nu})_\Gamma \,,
\nonumber\\
  \delta \mathcal{G}^+_{\mu\nu\Lambda} &=& - g\,\Lambda^P
  X_{PN\Lambda} \, \mathcal{G}^+_{\mu\nu}{}^N  -g \,
  \mathcal{N}_{\Lambda\Sigma}\,\Lambda^P X^\Gamma{}_P{}^\Sigma\,
  (\mathcal{G}^+_{\mu\nu} - \mathcal{H}^+_{\mu\nu})_\Gamma\,, 
\nonumber\\
  \delta(\mathcal{G}^+_{\mu\nu} - \mathcal{H}^+_{\mu\nu})_\Lambda
  &=&  g \, \Lambda^P(  X^\Gamma{}_{P\Lambda} -X^\Gamma{}_P{}^\Sigma
  \, \mathcal{N}_{\Sigma\Lambda})\,
  (\mathcal{G}^+_{\mu\nu} - \mathcal{H}^+_{\mu\nu})_\Gamma\,.
\end{eqnarray}
Hence $\delta\mathcal{G}_{\mu\nu}{}^M=-g\,\Lambda^PX_{PN}{}^M\,
\mathcal{G}_{\mu\nu}{}^N$, just as the variation of the abelian field
strengths $G_{\mu\nu}{}^M$ in the absence of charges, up to terms
proportional to $\Theta^{\Lambda\alpha}(\mathcal{G}_{\mu\nu}
-\mathcal{H}_{\mu\nu})_\Lambda$. According to \cite{dWST6}, the latter
terms represent a set of field equations. The last equation of
(\ref{eq:delta-G/H}) then expresses the well-known fact that under a
symmetry field equations transform into field equations. As a result
the gauge algebra on these tensors closes according to
(\ref{eq:gauge-commutators}), up to the the same field equation.

Having identified some of the field equations, it is easy to see how
the Lagrangian should be modified. First of all, we replace the
abelian field strengths $F_{\mu\nu}{}^\Lambda$ in the Lagrangian
(\ref{eq:quadratic-L}) by $\mathcal{H}_{\mu\nu}{}^\Lambda$, so that 
\begin{equation}
  \label{eq:def-cal-G}
  \mathcal{G}_{\mu\nu\,\Lambda} = \mathrm{i}\, 
  \varepsilon_{\mu\nu\rho\sigma}\,
  \frac{\partial\mathcal{L}_{\mathrm{vector}}}
  {\partial{\mathcal{H}}_{\rho\sigma}{}^\Lambda}   \;. 
\end{equation}
Under general variations of the vector and tensor fields we then
obtain the result,
\begin{equation}
  \label{eq:var-L-vector}
  e^{-1}\delta\mathcal{L}_{\mathrm{vector}} = -\mathrm{i}
  \mathcal{G}^{+\mu\nu}{}_\Lambda \Big[ D_\mu\delta A_\nu{}^\Lambda +
  \ft14 g\Theta^{\Lambda\alpha} (\delta B_{\mu\nu\alpha} - 2d_{\alpha
  PQ} A_\mu{}^P \delta A_\nu{}^Q)\Big ] + \mathrm{h.c.}  \,.
\end{equation}
From this expression the reader can check that the Lagrangian
(\ref{eq:quadratic-L}) is indeed invariant under the tensor gauge
transformations. Even when including the gauge transformations of the
matter fields, the Lagrangian is, however, not invariant under the
vector gauge transformations. For invariance it is necessary to
introduce the following universal terms to the Lagrangian
\cite{dWST6},
\begin{eqnarray}
  \label{eq:topological}
  {\cal L}_{\rm top} &=&
  \ft1{8}\mathrm{i} g\, \varepsilon^{\mu\nu\rho\sigma}\,
  \Theta^{\Lambda\alpha}\,B_{\mu\nu\,\alpha} \,
  \Big(2\partial_{\rho} A_{\sigma\,\Lambda} + g\,
  X_{MN\,\Lambda} \,A_\rho{}^M A_\sigma{}^N
  +\ft14 g\,\Theta_{\Lambda}{}^{\beta}B_{\rho\sigma\,\beta} \Big)
  \nonumber\\[.9ex]
  &&{}
  +\ft1{3}\mathrm{i} g\, \varepsilon^{\mu\nu\rho\sigma}X_{MN\,\Lambda}\,
  A_{\mu}{}^{M} A_{\nu}{}^{N}
  \Big(\partial_{\rho} A_{\sigma}{}^{\Lambda}
  +\ft14 g\,X_{PQ}{}^{\Lambda} A_{\rho}{}^{P}A_{\sigma}{}^{Q}\Big)
  \nonumber\\[.9ex]
  &&{}
  +\ft1{6}\mathrm{i} g\, \varepsilon^{\mu\nu\rho\sigma}X_{MN}{}^{\Lambda}\,
  A_{\mu}{}^{M} A_{\nu}{}^{N}
  \Big(\partial_{\rho} A_{\sigma}{}_{\Lambda}
  +\ft14 g\,X_{PQ\Lambda} A_{\rho}{}^{P}A_{\sigma}{}^{Q}\Big)\;.
\end{eqnarray}
The first term represents a topological coupling of the antisymmetric
tensor fields with the magnetic gauge fields, and the last two terms
are a generalization of the Chern-Simons-like terms
(\ref{eq:Chern-Simons}) that we encountered in the previous
subsection. Under variations of the vector and tensor fields, this
Lagrangian varies into (up to total derivative terms)
\begin{equation}
  \label{eq:var-L-top}
  e^{-1}\delta\mathcal{L}_{\mathrm{top}} = \mathrm{i}
  \mathcal{H}^{+\mu\nu\Lambda} \, D_\mu\delta A_{\nu\Lambda}  +
  \ft14 \mathrm{i} g\, \mathcal{H}^{+\mu\nu}{}_\Lambda
  \,\Theta^{\Lambda\alpha} (\delta B_{\mu\nu\alpha} - 2d_{\alpha 
  PQ} A_\mu{}^P \delta A_\nu{}^Q) + \mathrm{h.c.} \,.
\end{equation}
Under the tensor gauge transformations this variation becomes equal to
the real part of $2\mathrm{i} g\, \mathcal{H}^{+\mu\nu
  M}\,\Theta_M{}^\alpha\,D_\mu\Xi_{\nu\alpha}$. This expression equals
a total derivative by virtue of the invariance of the embedding
tensor, the constraint (\ref{eq:closure-constraint2}), and the Bianchi
identity
\begin{equation}
  \label{eq:Bianchi-D}
  D_{[\mu}{\cal H}_{\nu\rho]}{}^{M} = \ft13g\,Z^{M,\alpha}\,\Big[ 
  3\, D_{[\mu} B_{\nu\rho]\,\alpha} +6 \,d_{\alpha\, NP}\,A_{[\mu}{}^{N}
  (\partial_{\nu} A_{\rho]}{}^P+ \ft13 g X_{[RS]}{}^{P}
  A_{\nu}{}^{R}A_{\rho]}{}^{S})\Big]\,.
\end{equation}
In this Bianchi identity, $D_{\mu}{\cal H}_{\nu\rho}{}^{M} =
\partial_{\mu}{\cal H}_{\nu\rho}{}^{M} + gA_\mu{}^P X_{PN}{}^M
\mathcal{H}_{\nu\rho}{}^N$ and $D_\rho B_{\mu\nu\alpha}=
\partial_\rho B_{\mu\nu\alpha} - g A_\rho{}^M X_{M\alpha}{}^\beta
B_{\mu\nu\beta}$. This expression for the Bianchi identity is
suitable for our purpose here, but we note that it is not manifestly
covariant in this form, in view of the fact that the fully covariant
derivative of $\mathcal{H}_{\mu\nu}{}^M$ reads,
\begin{equation}
  \label{eq:new-cov-der-H}
  \mathcal{D}_\rho \mathcal{H}_{\mu\nu}{}^M =   
  \partial_\rho \mathcal{H}_{\mu\nu}{}^M + gA_\rho{}^P
  X_{PN}{}^M \,\mathcal{G}_{\mu\nu}{}^N  + gA_\rho{}^P
  X_{NP}{}^M \,
  (\mathcal{G}_{\mu\nu} - \mathcal{H}_{\mu\nu})^N \,,
\end{equation}
and the covariant field strength of the tensor fields equals
\begin{equation}
  \label{eq:tensor-H}
  {\cal H}_{\mu\nu\rho\,\alpha} \equiv 3\, D_{[\mu}
  B_{\nu\rho]\,\alpha} +6 \,d_{\alpha\, MN}\,A_{[\mu}{}^{M}
  \Big(\partial_{\nu} A_{\rho]}{}^N+ \ft13 g X_{[RS]}{}^{N}
  A_{\nu}{}^{R}A_{\rho]}{}^{S} + \mathcal{G}_{\nu\rho]}{}^N -
  \mathcal{H}_{\nu\rho]}{}^N \Big) \;. 
\end{equation}
The manifestly covariant form of the Bianchi identity
(\ref{eq:Bianchi-D}) then reads, 
\begin{equation}
  \label{eq:bianchi-H}
  \mathcal{D}_{[\mu}{\cal H}_{\nu\rho]}{}^{M} = 
  \ft13g\,Z^{M,\alpha}\,{\cal H}_{\mu\nu\rho\,\alpha} \;.  
\end{equation}

The various modifications described in this subsection ensure the
gauge invariance of the Lagrangian $\mathcal{L}_{\mathrm{vect}} +
\mathcal{L}_{\mathrm{top}}$, provided we include the gauge
transformations of the scalar fields \cite{dWST6}.
Furthermore, variation of the tensor fields yields the field equations
identified above,
\begin{equation}
  \label{eq:B-field-eq}
  \delta\mathcal{L}_{\mathrm{vector}} +
  \delta\mathcal{L}_{\mathrm{top}} = -  \ft14 \mathrm{i}g\,
  \, \delta B_{\mu\nu \alpha}\; 
  \Theta^{\Lambda\alpha} \Big[(\mathcal{G}^{+\mu\nu}
  -\mathcal{H}^{+\mu\nu})_\Lambda - (\mathcal{G}^{-\mu\nu}
  -\mathcal{H}^{-\mu\nu})_\Lambda \Big] \,.
\end{equation}
This result shows that the Lagrangian is invariant under variations of
the tensor fields for those components that are projected to zero by
the embedding tensor component $\Theta^{\Lambda\alpha}$. This implies
that these components of the tensor field do not appear in the action,
which plays a crucial role in ensuring that the number of degrees of
freedom will remain unchanged.

A similar phenomenon takes place for the magnetic gauge fields
$A_{\mu\Lambda}$. Evaluating the field equation for the gauge fields
$A_\mu{}^M$ one finds that the equation for the magnetic gauge
fields is only proportional to $\Theta^{\Lambda\alpha} \delta
A_{\mu\Lambda}$.  To see this, one evaluates
\begin{equation}
  \label{eq:A-field-eq}
  \delta\mathcal{L}_{\mathrm{vector}} +
  \delta\mathcal{L}_{\mathrm{top}} =  \ft12 \mathrm{i}\,
  \varepsilon^{\mu\nu\rho\sigma} \,D_\nu \mathcal{G}_{\rho\sigma}{}^M
  \Omega_{MN} \delta A_\mu{}^N \,,
\end{equation}
up to a total derivative and up to terms that vanish as a result of
the field equation for $B_{\mu\nu\alpha}$. Here one makes use of
(\ref{eq:lin}). Note that $D_\nu\mathcal{G}_{\rho\sigma}{}^M=
\mathcal{D}_\nu\mathcal{G}_{\rho\sigma}{}^M$, and furthermore that
$\mathcal{D}_\nu\mathcal{G}_{\rho\sigma}{}^\Lambda=
\mathcal{D}_\nu\mathcal{H}_{\rho\sigma}{}^\Lambda$, 
up to terms that
vanish by virtue of the field equation for $B_{\mu\nu\alpha}$.  Using
the Bianchi identity (\ref{eq:bianchi-H}) we can thus rewrite
(\ref{eq:A-field-eq}) as follows,
\begin{equation}
  \label{eq:A-field-eq-2}
  \delta\mathcal{L}_{\mathrm{vector}} +
  \delta\mathcal{L}_{\mathrm{top}} = \ft12 \mathrm{i}\,
  \varepsilon^{\mu\nu\rho\sigma} \left[- {D}_\nu
  \mathcal{G}_{\rho\sigma\Lambda} \,\delta A_\mu{}^\Lambda + \ft16 g\,
  \mathcal{H}_{\nu\rho\sigma\alpha} \,\Theta^{\Lambda\alpha} \delta
  A_{\mu\Lambda} \right] \,,
\end{equation}
under the same conditions as stated above.  Note that the minimal
coupling of the gauge fields is always proportional to the embedding
tensor. Therefore the full Lagrangian does not depend on those
components of the magnetic gauge fields that are projected to zero by
the embedding tensor component $\Theta^{\Lambda\alpha}$.

In the spririt of the analysis presented in \cite{deWit:2005hv}, one
may thus regard the absence of the components of $B_{\mu\nu\alpha}$ and
$A_{\mu\Lambda}$ as resulting from an additional gauge invariance
(which would then lead to rank-three tensors fields). However, since
these fields will not appear in the Lagrangian, there is no need for
doing so. Somewhat unexpectedly, and not in line with the general
analysis of the vector-tensor hierarchies, there is an additional
(local) invariance which involves only the tensor field \cite{dVdW}, 
\begin{equation}
  \label{eq:Delta-invariance}
  \Theta^{\Lambda\alpha}\delta B_{\mu\nu\alpha} \propto
  \Delta^{\Lambda\Sigma\rho}{}_\rho\,
  (\mathcal{G}-\mathcal{H})_{\mu\nu\Sigma} - 6\,
  \Delta^{(\Lambda\Sigma)\rho}{}_{[\rho}\,   
  (\mathcal{G}-\mathcal{H})_{\mu\nu]\Sigma} \,,
\end{equation}
where $\Delta^{\Lambda\Sigma\mu}{}_\nu= \Theta^{\Lambda\alpha}
\Delta_\alpha{}^{\Sigma\mu}{}_\nu$. This new invariance has, of
course, a role to play in balancing the degrees of freedom, but in
\cite{dWST6} this aspect was bypassed in the analysis. We note that
not all of these gauge invariances have a bearing on the dynamic modes
of the theory as they also act on fields that play an auxiliary role.

In spite of the modifications above, supersymmetry will be broken by
the gauging. In section \ref{sec:Lagr+transf} we show how
supersymmetry can be restored. But first we have to deal with the
effect of the gauge transformations on the scalar fields.

\section{The $T$-tensor}
\setcounter{equation}{0}
\label{sec:T-tensor}
We already stressed in the introduction that the scalar fields 
parametrize the ${\rm E}_{7(7)}/{\rm SU}(8)$ coset 
space.\footnote{
  Strictly speaking the isotropy group equals ${\rm SU}(8)/{\bf Z}_2$.
}  
These fields are described by a space-time dependent matrix ${\cal
  V}(x) \in {\rm E}_{7(7)}$ (taken in the fundamental ${\bf 56}$
representation) which transforms from the right under local ${\rm
  SU}(8)$ and from the left under rigid \Exc7. The matrix $\vv$ can be
used to elevate the embedding tensor to the so-called $T$-tensor,
which is the ${\rm SU}(8)$-covariant, field-dependent, tensor that
appears in the fermionic masslike terms and the scalar potential of
the Lagrangian.  The $T$-tensor is thus defined by, 
\begin{equation}
  \label{eq:T-M-alpha}
  T_{\underline M}{}^{\underline{\a}}[\Theta,\phi]\,t_{\underline\a} =
  {\cal V}^{-1}{}_{\!\!\!\!\underline{M}}{}^N\, \Theta_N{}^\a\, ({\cal
  V}^{-1}t_\a {\cal V} )\;,
\end{equation}
where the underlined indices refer to local ${\rm SU}(8)$. The
appropriate representation for (\ref{eq:T-M-alpha}) is the 
${\bf 56}$, so that we may write,
\begin{equation}
\label{eq:T-theta}
T_{\underline M \underline N}{}^{\underline P}[\Theta,\phi] = {\cal
  V}^{-1}{}_{\!\!\!\!\underline{M}}{}^M\; {\cal
  V}^{-1}{}_{\!\!\!\!\underline{N}}{}^N\; {\cal
  V}_P{}^{\underline{P}}\; X_{MN}{}^P\;.   
\end{equation}
Because the constraints on the embedding tensor are covariant under
\Exc7, it is clear that they induce a corresponding set of ${\rm
  SU}(8)$ covariant constraints on the $T$-tensor. 

However, we employ a somewhat unconventional definition of the
coset representative $\vv$. Note that the $T$-tensor is defined in an
${\rm SU}(8)$ covariant basis, where the maximal compact ${\rm SU}(8)$
subgroup of \Exc7 takes a block-diagonal form according to the
branching under ${\rm SU}(8)$, ${\bf 56} \to {\bf 28} + \overline{\bf
  28}$. This implies the existence of a pseudo-real vector
$U^{\underline M}$ decomposing according to $U^{\underline M} =
(U^{ij}, U_{kl})$, where $ij$ and $kl$ denote antisymmetric index
pairs with $i,j,k,l= 1,\ldots ,8$. This basis facilitates the coupling
to the fermions which transform under $\mathrm{SU}(8)$. On the other 
hand, just as in the preceding section,
we decompose the gauge fields in a real basis according to $V^M =
(V^\Lambda,V_\Sigma)$ which branches under the maximal real ${\rm
  SL}(8)$ subgroup of \Exc7 according to ${\bf 56} \to {\bf 28} +{\bf
  28}^\prime$. Therefore we define 56-dimensional complex vectors
$\vv_M{}^{ij}= (\vv_\Lambda{}^{ij}, \vv^{\Sigma \,ij})$ and their
complex conjugate $\vv_{M\,ij}=(\vv_{\Lambda\,ij},\vv^\Sigma{}_{ij})$,
which together constitute a $56\times 56$ matrix $\vv$,
\begin{equation}
  \label{eq:56-bein}
  \vv_M{}^{\underline N} =\Big(\vv_M{}^{ij}, \vv_M{}_{kl} \Big) = 
  \pmatrix{\vv_\Lambda{}^{ij}&\vv_{\Lambda\,kl}\cr
\noalign{\vskip 4mm}
       \vv^{\Sigma\,ij} & \vv^\Sigma{}_{kl}\cr}\,.
\end{equation}
This matrix thus transforms under rigid \Exc7 from the left and under
local ${\rm SU}(8)$ from the right. It does not really constitute an
element of \Exc7, but it is equal to a constant matrix (to account for
the different bases adopted on both sides) times a space-time
dependent element of \Exc7. We note the following useful properties of
$\vv_M{}^{\underline N}$, which also fix the
normalization,
\begin{eqnarray}
  \label{eq:VV-orthogonal}
  \vv_M{}^{ij} \,\vv_{N\,ij} - \vv_{M\,ij}\, \vv_N{}^{ij}  &=&  
  \mathrm{i}\,\Omega_{MN}\,, \nonumber\\ 
  \Omega^{MN} \,\vv_M{}^{ij} \,\vv_{N\,kl} &=& 
  \mathrm{i}\,\delta^{ij}{}_{kl}\,, \nonumber\\  
  \Omega^{MN} \,\vv_M{}^{ij} \, \vv_N{}^{kl} &=& 0\,. 
\end{eqnarray}
The sign on the right-hand side is determined by the relative phase
between $\vv_\Lambda{}^{ij}$ and $\vv^{\Lambda ij}$. Because we have
already fixed the definition of the \Exc7 transformations on the field
strenghts $(F_{\mu\nu}{}^\Lambda,G_{\mu\nu\Lambda})$, we can no longer
adjust this relative phase. Therefore we must distinguish two
different cases characterized by the sign on the right-hand side of
(\ref{eq:VV-orthogonal}). As it turns out, supersymmetry selects the
sign shown above. 

The equations (\ref{eq:56-bein}) and (\ref{eq:VV-orthogonal}) imply
that the inverse coset representative $\vv^{-1}$ reads,
\begin{equation}
  \label{eq:inverse-56-bein}
  [\,\vv^{-1}]_{\underline M}{}^{N} = \mathrm{i}\Omega^{NP}
  \,\Big(- \vv_{P\,ij}, \vv_P{}^{kl} \Big) =   
  \pmatrix{- \mathrm{i} \vv^\Lambda{}_{ij} & \mathrm{i} \vv_{\Sigma\,ij}\cr
\noalign{\vskip 4mm}
      \mathrm{i}\vv^{\Lambda\,kl} & -\mathrm{i} \vv_\Sigma{}^{kl}\cr}\,.
\end{equation}
The most relevant restriction is, however, not captured by
(\ref{eq:VV-orthogonal}), namely that $\vv_M{}^{\underline N}$ can be
written as a constant tensor ${\stackrel{\circ}{\vv}}_M{}^{Z}$ times a
space-time dependent \Exc7 matrix $\vv_{Z}{}^{\underline N}(x)$. The
latter $56\times 56$ matrix, sometimes called the 56-bein, is usually
expressed in the form,
\begin{equation}
  \label{eq:u-v}
  \vv_{Z}{}^{\underline N}(x) 
  = \pmatrix{ u^{ij}{}_{\!IJ}(x) & -v_{ kl  IJ}(x)\cr
    \noalign{\vskip 6mm}
    -v^{ijKL}(x) &   u_{kl}{}^{\!KL}(x)\cr} \,.
\end{equation}
The indices $I,J,\ldots$ and $i,j,\ldots$ take the values
$1,\ldots,8$, so that there are 28 antisymmetrized index pairs
representing the matrix indices of $\vv$; the row indices are ${Z} =
([IJ],[KL])$, and the column indices are ${\underline N}=
([ij],[kl])$, so as to remain consistent with the conventions of
\cite{deWitNic}. The above matrix is pseudoreal and belongs to ${\rm
  E}_{7(7)}\subset {\rm Sp}(56;\mathbb{R})$ in the fundamental
representation.  We use the convention where $u^{ij}{}_{\!IJ}=
(u_{ij}{}^{\!IJ})^\ast$ and $v_{ijIJ}=(v^{ij IJ})^\ast$.  The indices
$i,j,\ldots$ refer to local ${\rm SU}(8)$ transformations and capital
indices $I,J,\ldots$ are subject to rigid \Exc7 transformations.

A crucial question regards the nature of the constant matrix
$\stackrel{\circ}{\vv}$. Obviously (\ref{eq:VV-orthogonal}) leaves the
freedom to perform a redefinition by acting with an ${\rm
  Sp}(56;\mathbb{R})$ transformation from the left. Because $\vv_M$ is
defined with a lower index, such a transformation acts as follows,
\begin{eqnarray}
  \label{eq:V-em-duality}
  \vv_\Lambda{}^{ij} &\to& V_\Lambda{}^\Sigma \,\vv_\Sigma{}^{ij}
  -W_{\Lambda\Sigma}\, \vv^{\Sigma ij} \,,
  \nonumber\\ 
  \vv^{\Lambda ij} &\to& U^\Lambda{}_\Sigma \,\vv^{\Sigma ij} -
  Z^{\Lambda\Sigma}\, \vv_\Sigma{}^{ij}\,. 
\end{eqnarray}
These redefinitions lead to an obvious ambiguity in the definition of
$\stackrel{\circ}{\vv}$ and correspondingly in the definition of
$\vv_\Lambda{}^{ij}$ and $\vv^{\Lambda ij}$.  However, some of this
ambiguity can be removed, either by absorbing an \Exc7 transformation
emerging on the right into the definition of $\vv_{\underline
  P}{}^{Z}(x)$, or by absorbing an $\mathrm{GL}(28)$ transformation
emerging on the left into the definition of the gauge fields. The
ambiguity thus takes the form of an ${\rm E}_{7(7)}\backslash {\rm
  Sp}(56;\mathbb{R}) /{\rm GL}(28)$ matrix (or rather its inverse)
\cite{deWit02,dWST1}. The Lagrangian will implicitly depend on this
matrix, as it will be written in terms of $\vv_\Lambda{}^{ij}$ and
$\vv^{\Lambda ij}$.

Let us now briefly discuss the pseudoreal representation of \Exc7. The
maximal compact subgroup $\mathrm{SU}(8)$ coincides with the
R-symmetry group of four-dimensional $N=8$ supersymmetry which is
relevant for the fermions, as the chiral and antichiral gravitino and
spinor fields transform in the ${\bf 8}+\overline{\bf 8}$, and ${\bf
  56}+ \overline{\bf 56}$ representation of that group. Therefore the
pseudoreal basis, based on the $\mathrm{SU}(8)$ decomposition ${\bf
  56} \to \overline{\bf 28} + {\bf 28}$, is particularly relevant. In
the ${\bf 56}$ representation, the basis vectors in the ${\bf 56}$
representation are then denoted by $(z_{IJ},z^{KL})$ with $z^{IJ} =
(z_{IJ})^\ast$; here the indices are antisymmetrized index pairs
$[IJ]$ and $[KL]$ and $I,J,K,L = 1,\ldots,8$. The $z^{IJ}$ transform
according to the ${\bf 28}$ representation of ${\rm SU}(8)$.
Infinitesimal ${\rm Sp}(56;\mathbb{R})$ transformations now take the
form,
\begin{eqnarray}
  \label{eq:delta-z}
\d z_{IJ} &=& \Lambda_{IJ}{}^{\!KL} \,z_{KL} + \Sigma_{IJKL}
\,z^{KL}\,,\nn\\
\d z^{IJ} &=& \Lambda^{IJ}{}_{\!KL} \,z^{KL} + \Sigma^{IJKL}
\,z_{KL}\,,
\end{eqnarray}
where $\Lambda_{IJ}{}^{\!KL}$ and $\Sigma_{IJKL}$ are subject to the
conditions 
\begin{eqnarray}
  \label{eq:sp56-properties}
(\Lambda_{IJ}{}^{\!KL})^\ast =  \Lambda^{IJ}{}_{\!KL}=
-\Lambda_{KL}{}^{\!IJ} \,, \qquad
   (\Sigma_{IJKL})^\ast =  \Sigma^{KLIJ} \,.
\end{eqnarray}
The matrices $\Lambda_{IJ}{}^{KL}$ are associated with the maximal
compact ${\rm U}(28)$ subgroup.  In this basis the invariant
skew-symmetric tensor $\Omega$ is proportional to
\eqn{eq:symplectic-omega}. The ${\rm E}_{7(7)}$ subgroup of ${\rm
  Sp}(56;\mathbb{R})$ is obtained for fully antisymmetric
$\Sigma^{IJKL}$ with the additional restrictions,
\begin{eqnarray}
  \label{eq:E7}
&& \Lambda_{IJ}{}^{\!KL} = \d_{[I}{}^{[K} \,\Lambda_{J]}{}^{\!L]}  \,, 
\qquad \Lambda_I{}^J =    -\Lambda^J{}_I\,, \nn\\
&&
\Lambda_I{}^I = 0\,,\qquad \Sigma_{IJKL} = \ft1{24}
\varepsilon_{IJKLMNPQ} \,\Sigma^{MNPQ}\,.
\end{eqnarray}
The $\Lambda_{I}{}^{\!J}$ generate the group ${\rm SU}(8)$. 
Closure of the full algebra is ensured by the fact that two tensors $\Sigma_1$
and $\Sigma_2$ satisfy the relation
\begin{eqnarray}
  \label{eq:sigma-algebra}
  &&\Sigma_{1\,IJMN} \,\Sigma_2{}^{MNKL} - \Sigma_{2\,IJMN}
  \,\Sigma_1{}^{MNKL} \nn\\ 
  && = \ft23\,\d_{[I}{}^{[K} \,\Big(\Sigma_{1\,J]MNP}
  \,\Sigma_2{}^{L]MNP}-\Sigma_{2\,J]MNP} \,\Sigma_1{}^{L]MNP}\Big) \,,
\end{eqnarray}
which follows from the selfduality of $\Sigma$.  All this is in accord with
the branching of the adjoint representation of \Exc7 with respect to its ${\rm
  SU}(8)$ subgroup:  ${\bf 133}\to {\bf 63} + {\bf 70}$.

Before returning to the $T$-tensor, let us first reconsider the
representation of the scalar fields based on $\vv_\Lambda{}^{ij}$ and
$\vv^{\Lambda ij}$. Under arbitrary variations of the \Exc7 matrix
(\ref{eq:u-v}) we note the result, 
\begin{equation}
  \label{eq:cartan-form}
  [\vv^{-1}]_{\underline M}{}^N \,\delta\vv_N{}^{\underline P} = 
     [\vv^{-1}]_{\underline M}{}^{Z}
   \;\delta\vv_{Z}{}^{\underline P}  \,,
\end{equation}
which follows from the fact that the constant matrix
$\stackrel{\circ}{\vv}$ cancels in the expression on the left-hand
side. This observation leads to
\begin{eqnarray}
  \label{eq:V-dV-1}
  \vv_{M ij} \,\delta\vv_N{}^{kl} \,\Omega^{MN} &=& - \mathrm{i}\Big(
  u_{ij}{}^{IJ} \,\delta 
  u^{kl}{}_{IJ} - v_{ij IJ} \,\delta   v^{kl IJ} \Big) \;,\nn \\
  \vv_{M ij} \,\delta\vv_{N kl} \,\Omega^{MN} &=& - \mathrm{i}\Big(
  v_{ij IJ} \,\delta   u_{kl}{}^{IJ} -
  u_{ij}{}^{IJ} \,\delta v_{kl IJ}  \Big) \;.
\end{eqnarray}
The expression on the right-hand side shows that the equation
(\ref{eq:cartan-form}) can be decomposed into the generators of \Exc7.
The first term should be proportional to the $\mathrm{SU}(8)$
generators in the ${\bf 28}$ representation, and the second term
should belong to the ${\bf 70}$ representation. Using these
restrictions, we derive,
\begin{eqnarray}
  \label{eq:V-dV}
  \vv_{M ij} \,\delta\vv_N{}^{kl} \,\Omega^{MN}   &=&{}
  \ft{3}{2}\delta_{[i}{}^{[k} \; \vv_{M j]m} \,\delta\vv_N{}^{l]m}
  \,\Omega^{MN}  \,, \nn\\
  \vv_{M ij} \,\delta\vv_{N kl} \,\Omega^{MN}   &=&{}
  \vv_{M [ij} \,\delta\vv_{N kl]} \,\Omega^{MN}  \,, \nn\\
  \vv_{M}{}^{ij} \,\delta\vv_N{}^{kl} \,\Omega^{MN} &=& - \ft1{24}
  \varepsilon^{ijklmnpq} \,   \vv_{M mn} \,\delta\vv_{N pq}
  \,\Omega^{MN}  \,. 
\end{eqnarray}
In what follows these equations play an important role. 

Let us now return to the $T$-tensor. First we draw attention to the
fact that, when treating the embedding tensor as a spurionic object
that transforms under the duality group, the equations of motion, the
Bianchi identities and the transformation rules remain formally
invariant under \Exc7. Under the latter $\Theta_M{}^\alpha$ would
transform as $\Theta_M{}^\a\, t_\a \to
g_M{}^N\,\Theta_N{}^\a\,(g\,t_\a g^{-1})$, with $g \in {\rm
  E}_{7(7)}$. The same observation applies to the $T$-tensor. To make
this more explicit we note that {\it every} variation of the coset
representative can be expressed as a (possibly field-dependent) \Exc7
transformation acting on ${\cal V}$ from the right. For example, a
rigid \Exc7 transformation acting from the left can be rewritten as a
field-dependent transformation from the right,
\begin{equation}
  \label{eq:variation-V}
  {\cal V}\to {\cal
  V}^\prime = g\,{\cal V}= {\cal V} \, \sigma^{-1}\,,
\end{equation}
with $\sigma^{-1}= {\cal V}^{-1}\,g\,{\cal V}\in {\rm E}_{7(7)}$, but also
a supersymmetry transformation can be written in this form.
Consequently, these variations of ${\cal V}$ induce the following
transformation of the $T$-tensor, 
\begin{equation}
  \label{eq:T-linear}
  T_{\underline M \underline N}{}^{\underline P}
  \to T^{\prime}_{\underline M\underline N}{}^{\underline P} =
  \sigma_{\underline M}{}^{\underline Q}\,
  \sigma_{\underline N}{}^{\underline R}\,
  (\sigma^{-1})_{\underline S}{}^{\underline P}\;
  T_{\underline Q\underline R}{}^{\underline S} \;. 
\end{equation}
This implies that the $T$-tensor constitutes a representation of
\Exc7. Observe that this is {\it not} an invariance statement; rather it means
that the $T$-tensor (irrespective of the choice for the corresponding
embedding tensor) varies under supersymmetry or any other transformation in a
way that can be written as a (possibly field-dependent)
\Exc7-transformation. Note also that the transformation assignment of 
the embedding tensor and the $T$-tensor are opposite in view of the
relationship between $g$ and $\sigma$, something that is important in
practical applications.  

Subsequently we determine the $T$-tensor according to
(\ref{eq:T-theta}). First we define
\begin{eqnarray}
  \label{eq:P-Q-gauged}
 \Omega^{NP} \,\vv_{N\,ij} X_{MP}{}^Q \,\vv_Q{}^{kl}  &=&
  {} -\mathrm{i} {\cal Q}_{M\, ij}{}^{\!kl} \;,\nonumber \\
 \Omega^{NP} \,\vv_{N\,ij} X_{MP}{}^Q \,\vv_{Q kl}  &=&
   {} - \mathrm{i} {\cal P}_{M\, ijkl}\;.
\end{eqnarray}
We note that $\mathcal{Q}_M$ and $\mathcal{P}_M$ are subject to
constraints,
\begin{equation}
  \label{eq:Z-Q-P}
  Z^{M,\alpha} \,\mathcal{Q}_{Mij}{}^{kl} = 0\;, \qquad Z^{M,\alpha}
  \,\mathcal{P}_{Mijkl} = 0\; ,
\end{equation}
by virtue of the quadratic constraint (\ref{eq:closure-constraint2}).
The tensor $Z^{M,\alpha}$ was defined in (\ref{eq:def-Z}). For the
convenience of the reader, we also note the relation,
\begin{equation}
  \label{eq:V-Q-P}
  X_{MN}{}^P\,\vv_P{}^{ij} = \mathcal{P}_M{}^{ijkl}\,\vv_{Nkl}
  +\mathcal{Q}_{Mkl}{}^{ij} \,\vv_N{}^{kl} \,. 
\end{equation}
The generators $X_M$ define a subgroup of \Exc7 in a certain
electric/magnetic duality basis, which in (\ref{eq:P-Q-gauged}) is
converted to the pseudoreal representation.  Compatibility with the
Lie algebra of {\rm \Exc7} implies that ${\cal P}_M{}^{ijkl}$ is a
selfdual ${\rm SU}(8)$ tensor, \be {\cal P}_M{}^{ijkl} =
\ft1{24}\,\varepsilon^{ijklmnpq}\, {\cal P}_{M\,mnpq}\,, \ee and that
${\cal Q}_M$ transforms as a connection associated with ${\rm SU}(8)$.
Hence, ${\cal Q}_{M \,ij}{}^{\!kl}$ satisfies the decomposition,
\begin{equation}
{\cal Q}_{M\, ij}{}^{\!kl}= \delta_{[i}{}^{[k}\, {\cal
  Q}_{M\,j]}{}^{\!l]}\,, 
\end{equation}
with ${\cal Q}_{M}{}^{\!i}{}_{\!j} = - {\cal Q}_{M j}{}^i$ and
${\cal Q}_{M i}{}^i=0$. Decomposing 
\begin{equation}
T_{\underline M \underline N}{}^{\underline P} = \Big(T_{ij \underline
  N}{}^{\underline P}, T^{kl}{}_{\underline N}{}^{\underline
  P}\Big)\;, 
\end{equation}
we write the components of the $T$-tensor in matrix notation, 
\begin{equation}
  \label{eq:TE7.1}
  T_{ij}=  \pmatrix{- \ft23\d_{[k}{}^{[p}\, T^{q]}{}_{l]ij} & 
    {1\over 24}\varepsilon_{klrstuvw} \, T^{tuvw}{}_{ij} \cr
    \noalign{\vskip5mm}
      T^{mnpq}{}_{ij} &
             \ft23 \d_{[r}{}^{[m} \,T^{n]}{}_{s]ij}\cr}\;,
\end{equation} 
where $({}_{[kl]},{}^{[mn]})$ are the row indices and
$({}^{[pq]},{}_{[rs]})$ the column indices, and
\begin{equation}
  \label{eq:TE7.2}
  T^{ij}= \pmatrix{\ft23\d_{[k}{}^{[p} \,T_{l]}{}^{q]ij} &
  T_{klrs}{}^{ij} \cr \noalign{\vskip5mm} {1\over
    24}\varepsilon^{mnpqtuvw} T_{tuvw}{}^{ij} & -
  \ft23\d_{[r}{}^{[m}\, T_{s]}{}^{n]ij} \cr} \;.   
\end{equation}
Multiplicative factors have been included to make contact with the
definitions of \cite{deWitNic,deWit02,dWST1}.  In order to belong to
the Lie algebra of \Exc7, the matrix blocks in the above expressions
satisfy $T_k{}^{kij} = 0$ and $T^{klmn}{}_{ij} =T^{[klmn]}{}_{ij}$.
Note that we always use the convention where complex conjugation is
effected by raising and lowering of indices $\mathrm{SU}(8)$. 

Comparing the above expressions, one can directly establish the
following expressions,\footnote{
  Unlike in the original definition~(\ref{eq:T-theta}) the $\vv_M$ are
  only proportional to an \Exc7 group element, so that the
  proportionality factor in (\ref{eq:T1-tensor}) is not intrinsically
  defined. Our choice for this factor is such that our results remain
  as closely related as possible to the   original expressions of
  \cite{deWitNic}. }
\begin{eqnarray}
  \label{eq:T1-tensor}
  T_k{}^{lij} &=& {} \ft34 \mathrm{i}\,\Omega^{MN} 
  \,{\cal Q}_{M\,k}{}^l\,\vv_N{}^{ij}  \,, \nn \\
  T_{klmn}{}^{ij} &=&{} \ft12 \mathrm{i}\,\Omega^{MN} 
  \,{\cal P}_{M\,klmn} \,\vv_N{}^{ij} \,. 
\end{eqnarray}
Note that so far no constraints have been imposed on the $T$-tensor.

We already noted that every variation of the coset representative can
be cast in the form of an \Exc7 transformation acting on the right of
$\vv$. This implies that any variation of the $T$-tensor is again
proportional to the $T$-tensor itself ({\it c.f.}  \eqn{eq:T-linear}).
In view of the covariance under the ${\rm SU}(8)$ subgroup, the only
relevant variation is therefore
\begin{equation}
  \label{eq:delta-V}
  \vv\to \vv
  \pmatrix{ 0&\overline\Sigma\cr\noalign{\vskip1mm} \Sigma&0\cr} \,.
\end{equation}
In this way one can derive, 
\begin{eqnarray}
  \label{eq:T-variations}
  \delta T_i{}^{jkl} &=&  \Sigma^{jmnp} \,T_{imnp}{}^{kl} -\ft1{24}
  \varepsilon^{jmnpqrst} \,\Sigma_{imnp} \,T_{qrst}{}^{kl} +
  \Sigma^{klmn}\, T^j{}_{imn}  \nn\\
  &=& 2\, \Sigma^{jmnp}\, T_{imnp}{}^{kl} -\ft14\,\d_i{}^j\, \Sigma^{mnpq}\, 
  T_{mnpq}{}^{kl} + \Sigma^{klmn}\, T^j{}_{imn} \,, \nn\\
  \d T_{ijkl}{}^{mn} &=& {}-\ft43 \Sigma_{p[ijk}^{~}\, T_{l]}{}^{pmn}
  -\ft1{24} \varepsilon_{ijklpqrs}\, \Sigma^{mntu}_{~}\, T^{pqrs}{}_{tu}
  \,.   
\end{eqnarray}
This formula can be used for evaluating, for instance, space-time
derivatives or supersymmetry variations of the $T$-tensor, where one
must choose the appropriate expressions for $\Sigma, \overline
\Sigma\propto \vv^{-1}\,\delta\vv$.

Armed with these results we can now proceed and derive the constraints
on the $T$-tensor induced by the embedding tensor constraints
discussed in the previous section. First of all, as a consequence of
(\ref{eq:repres-constraint}), the $T$-tensor is constrained to the
${\bf 912}$ representation of \Exc7, which decomposes into a ${\bf
  36}$ and a ${\bf 420}$ representation of ${\rm SU}(8)$. This shows
that there must be a proportionality relation between
$T^{klmn}{}_{ij}$ and $\d_{[i}{}^{[k} \,T_{j]}{}^{lmn]}$, as both
sides can only contain the ${\bf 420}$ representation. Checking the
consistency of this with \eqn{eq:T-variations}, it follows
that
\begin{eqnarray}
  \label{eq:su8-repre-constraint}
  T^{klmn}{}_{ij} &=& -\ft43 \d_{[i}{}^{[k} \,T_{j]}{}^{lmn]}\,,
  \nonumber \\    
  T_i{}^{jkl} &=& -\ft34 \,A_{2i}{}^{jkl} - \ft32
  A_1{}^{j[k}\,\d^{l]}{}_i \,,   
\end{eqnarray}
where $A_{2i}{}^{jkl} = A_{2i}{}^{[jkl]}$, $A_{2i}{}^{jki}=0$ and
$A_1^{[ij]}=0$, so that $T_i{}^{[ijk]}=0$. Clearly $A_1$ and $A_2$
represent the ${\bf 36}$ and ${\bf 420}$ representations of ${\rm
  SU}(8)$, respectively.  These results are not new and were first
given in \cite{deWitNic},
but we prefer to give a self-contained derivation here to demonstrate
how to cast the group-theoretical restrictions into the equations that
one needs for the Lagrangian.  The ${\rm SU}(8)$ tensors $A_1$ and
$A_2$ appear in the Lagrangian in the masslike terms and in the scalar
potential that we will present in the next section. In fact, the
supersymmetry of the action to first order of the gauge coupling
constant $g$, depends crucially on \eqn{eq:su8-repre-constraint}. Note
that none of these results depend on the actual gauge group. The only
requirement is that the embedding tensor satisfies the constraints
discussed in the previous section.

We now turn to a discussion of the constraints that are quadratic in
the $T$-tensor. These constraints are sufficient for proving the
supersymmetry of the action to second order in $g$. In section~2 we
presented two alternative expressions for the quadratic constraint.
One is \eqn{eq:closure-constraint2}, which can be rewritten as an
equation for the $T$-tensor after suitable multiplication with $\vv$.
The results, which coincide with the ones derived in
\cite{deWitNic,dWST1}, take the form, 
\begin{eqnarray}
  \label{eq:quadratic}
T^k{}_{lij} \,T_n{}^{mij} - T_l{}^{kij} \, T^m{}_{nij}&=&0  \,,\nonumber \\
T^k{}_{lij} \,T_{mnpq}{}^{ij} +\ft1{24}\varepsilon_{mnpqrstu}\,
T_l{}^{kij}
\, T^{rstu}{}_{ij}&=& 0  \,,\nonumber \\
T_{irst}{}^{vw} \, T^{jrst}{}_{vw} - \ft18 \d_i^j\, T_{rstu}{}^{vw} \,
T^{rstu}{}_{vw}&=& 0 \,,\nonumber \\
T_{ijkr}{}^{vw} \,T^{mnpr}{}_{vw} - \ft94 \d_{[\,i}^{[m} \,
T_{jk]rs}{}^{vw} \,T^{np]rs}{}_{vw} +\ft1{16} \d{}_{\,i}^m{}_j^n{}_k^p
\,T_{rstu}{}^{vw} \, T^{rstu}{}_{vw} &=& 0 \,,   
\end{eqnarray}
where in the last
identity the antisymmetrization does not include the indices $v,w$.
Substituting the results of \eqn{eq:su8-repre-constraint}, these
equations reduce to,
\begin{eqnarray}
  \label{eq:quadratic2}
  A_{2}{}^k{}_{lij} \,A_{2n}{}^{mij} - A_{2l}{}^{kij} \, A_{2}{}^m{}_{nij}
  -4A_{2}{}^{(k}{}_{lni}A_{1}^{m)i}-4A_{2(n}{}^{mki}A_{1l)i}&& \nn\\
  {}-2\delta_{l}^{m}A_{1ni}A_{1}{}^{ki}+2\delta_{n}^{k}A_{1li}A_{1}{}^{mi}
  &=&0\,,\nonumber \\
  A_{2}{}^i{}_{jk[m} \,A_{2}{}^k{}_{npq]}
  +A_{1jk}\delta^{i}_{[m}A_{2}{}^{k}{}_{npq]}
  -A_{1j[m}A_{2}{}^{i}{}_{npq]} &&\nn\\
  {}+\ft1{24}\varepsilon_{mnpqrstu}\,
  (A_{2j}{}^{ikr}\, A_{2k}{}^{stu}
  +A_{1}^{jk}\delta_{j}^{r}A_{2k}{}^{stu}-A_{1}^{ir}A_{2j}{}^{stu})&=& 0  \,,
  \nonumber\\
9\, A_2{}^m{}_{ikl} \, A_{2m}{}^{jkl} - A_2{}^j{}_{klm}\,A_{2i}{}^{klm} -
\d_i{}^j\, 
A_2{}^n{}_{klm} \, A_{2n}{}^{klm}&=& 0 \,,\nonumber \\
A_2{}^r{}_{ijk} \,A_{2r}{}^{mnp} - 9\,A_2{}^{[m}{}_{r[ij}\,
A_{2k]}{}^{np]r}
- 9\, \d_{[i}{}^{[m} \, A_2{}^n{}_{rsj}\,A_{2k]}{}^{p]rs}&& \nn\\
- 9\, \d_{[i\,j}{}^{[mn}\,A_2{}^u{}_{k]rs}\, A_{2u}{}^{p]rs}
+\d{}_{\,i}^m{}_j^n{}_k^p \,A_2{}^u{}_{rst} \, A_{2u}{}^{rst} &=& 0 \,,
\end{eqnarray}
where the antisymmetrizations in the last equation apply to the index
triples $[ijk]$ and $[mnp]$. Note that the representation content of
these four constraint equations is ${\bf 945} + \overline{\bf 945}
+{\bf 63}$, ${\bf 3584} + {\bf 378}+\overline{\bf 378} +{\bf 70}$,
${\bf 63}$ and  ${\bf 2352}$, respectively. 

As we intend to demonstrate in the following, consistent gaugings are
characterized by embedding tensors that satisfy two constraints
(\ref{eq:repres-constraint}) and (\ref{eq:closure-constraint2}), one
linear and one quadratic in this tensor. These two constraints lead to
corresponding constraints on the $T$-tensor, namely 
(\ref{eq:su8-repre-constraint}) and (\ref{eq:quadratic2}).

\section{The Lagrangian and transformation rules}
\setcounter{equation}{0}
\label{sec:Lagr+transf}
In principle the Lagrangian and transformation rules are known from
\cite{deWitNic}, but we have to convert to the unconventional
definition of the coset representative. Furthermore we have to make
contact with the formalism of \cite{dWST6} to incorporate possible
magnetic charges.  The reader who wishes to avoid the complications
associated with the magnetic charges, can simply assume that an
appropriate electric/magnetic duality transformation has been
performed so that there are only electric charges (implying that
$\Theta^{\Lambda \alpha}=0$). But as we have indicated previously,
there is a variety of reasons why it is advantageous to remain in a
more general electric/magnetic duality frame.

\subsection{Coset geometry}
\label{sec:coset-geometry}
The first issue that we have to address is related to the coset
representative of ${\rm E}_{7(7)} /{\rm SU}(8)$. In particular we have
to write the composite ${\rm SU}(8)$ gauge fields ${\cal Q}_\mu$ and
the tensor ${\cal P}_\mu$ appearing in the kinetic term for the scalar
fields in terms of the $\vv_M{}^{ij}$. This proceeds in the standard
way. We assume the presence of 56 gauge fields $A_\mu{}^M$ which
couple to the charges $X_M$ as in (\ref{eq:cov-der}). The covariant
derivative,
\begin{equation}
  \label{eq:D-V}
  \mathcal{D}_\mu\vv_M{}^{ij} = \partial_\mu\vv_M{}^{ij}
  -\mathcal{Q}_{\mu \,kl}{}^{ij} \,\vv_M{}^{kl} - g\, A_\mu{}^P
  X_{PM}{}^N \,\vv_N{}^{ij}\,,
\end{equation}
is covariant with respect to $\mathrm{SU}(8)$, with corresponding connection 
\begin{equation}
  \label{eq:Q-su8}
{\cal Q}_{\m\, ij}{}^{\!kl}= \d_{[i}{}^{[k} \, {\cal Q}_{\m\,
j]}{}^{\!l]}\,,
\end{equation}
with ${\cal Q}_{\mu}{}^{\!i}{}_{\!j} = - {\cal Q}_{\m j}{}^i$ and
${\cal Q}_{\m i}{}^i=0$. Furthermore it is covariant under the
optional gauge transformations with generators $X_M$ and connections
$A_\mu{}^M$.  The $\mathrm{SU}(8)$ connection is, however, not an independent
field and determined by the condition,
\begin{equation}
  \label{eq:def-Q}
  \Omega^{MN} \,\vv_{M ij} \, \mathcal{D}_\m \vv_N{}^{kl}  = 0 \,,
\end{equation}
which yields
\begin{equation}
  \label{eq:expression-Q}
  {\cal Q}_{\mu \,i}{}^{j} =  \ft23 \mathrm{i} ( \vv_{\Lambda\,ik}
  \,\partial_\mu \vv^{\Lambda\,jk} - \vv^\Lambda{}_{ik}
  \,\partial_\mu \vv_\Lambda{}^{\!jk})    - g\, A_\mu{}^M\, 
  {\cal Q}_{M\,i}{}^{j}   \,, 
\end{equation}
where ${\cal Q}_{M\,i}{}^{j}$ is defined by (\ref{eq:P-Q-gauged}).

In addition we define an $\mathrm{SU}(8)$ tensor ${\cal P}_{\mu\,
  ijkl}$ which is invariant under the optional gauge group
$\mathrm{G}_g$,
\begin{equation}
  \label{eq:def-P}
  {\cal P}_{\mu\, ijkl} =  \mathrm{i}  \Omega^{MN} \,\vv_{M ij} \,
  \mathcal{D}_\m \vv_{N kl}=  \mathrm{i} ( \vv_{\Lambda\,ij}
  \,\mathcal{D}_\mu \vv^{\Lambda}{}_{kl}- \vv^\Lambda{}_{ij}
  \,\mathcal{D}_\mu \vv_{\Lambda\,kl}) \;,
\end{equation}
where the gauge fields contribute through the covariant derivative,
leading to $- g\, A_\mu{}^M\, {\cal P}_{M\,ijkl}$.  Compatibility with
the Lie algebra of {\rm \Exc7} implies that ${\cal P}_{\mu\,ijkl}$ is
a selfdual ${\rm SU}(8)$ tensor,
\begin{equation}
  \label{eq:P-selfdual}
 {\cal P}_\mu{}^{ijkl} =
\ft1{24}\,\varepsilon^{ijklmnpq}\, {\cal P}_{\m\,mnpq}\,.   
\end{equation}
Furthermore we note the useful identity,
\begin{equation}
  \label{eq:DV=P}
  \mathcal{D}_\mu \vv_M{}^{ij}=  \mathcal{P}_\mu{}^{ijkl} \,\vv_{M kl} 
  \,. 
\end{equation}

Applying a second derivative to (\ref{eq:def-Q})~(\ref{eq:DV=P}) leads to 
integrability conditions known as the Cartan-Maurer equations, 
\begin{eqnarray}
  \label{eq:GECM-Q-P}
  F_{\mu\nu}({\cal Q})_i{}^j & = &  -\ft43\,{\cal P}_{[\mu}{}^{\!jklm}
  \, {\cal P}_{\nu]iklm} -g\,\mathcal{F}_{\m\n}{}^{M} \,{\cal
    Q}_{M\,i}{}^j\,, 
  \nonumber \\
   D_{[ \mu}{\cal P}_{\nu]}{}^{ijkl} &=& - \ft12
   g\,\mathcal{F}_{\m\n}{}^{M} \,{\cal P}_{M}{}^{ijkl} \,,   
\end{eqnarray}
where ${\cal Q}_{M\,i}{}^{j}$ and $\mathcal{P}_M{}^{ijkl}$ are defined
by (\ref{eq:P-Q-gauged}), 
\begin{equation}
  \label{eq:su8-field-strength}
 F(\mathcal{Q})_{\mu\nu i}{}^j = \pa_\mu \mathcal{Q}_{\nu i}{}^j  -
\pa_\nu \mathcal{Q}_{\mu i}{}^j   + \mathcal{Q}_{[\mu i}{}^k\,
\mathcal{Q}_{\nu]k}{}^j  \,, 
\end{equation}
is the $\mathrm{SU}(8)$ field strength, and $\mathcal{F}_{\mu\nu}{}^M$
was already defined in (\ref{eq:field-strength}). 
These Cartan-Maurer equations are important for deriving the
supersymmetry of the action. The order-$g$ terms violate the
supersymmetry of the original ungauged Lagrangian as they induce new
supersymmetry variations of the gravitino kinetic terms and the
Noether term, which are proportional to the field strengths
$\mathcal{F}_{\m\n}{}^{M}$ and also to the $T$-tensor.  

\subsection{The ungauged Lagrangian}
\label{sec:ungauged-lagrangian}
In this subsection we briefly introduce the ungauged Lagrangian of
$N=8$ supergravity in the notation of this paper. Up to terms
proportional to the field equations of the gauge fields, this
Lagrangian is invariant under an \Exc7 subgroup of the
$\mathrm{Sp}(56,\mathbb{R})$ electric/magnetic duality group. The
most crucial part of the Lagrangian concerns the 28 electric vector
fields $A_\mu{}^\Lambda$ (their magnetic duals $A_{\mu\Lambda}$ are
absent as we already discussed in subsection \ref{sec:special-basis}),
which are only invariant under a subgroup of \Exc7. The field
equations for these vector fields and the Bianchi identities for their
field strengths constitute 56 equations, given in
(\ref{eq:eom-bianchi}), which are subject to electric/magnetic duality
transformations. Only the vector field strengths $F_{\mu\nu}{}^M$ and
the scalar fields contained in $\vv_M{}^{ij}$ are subject to the \Exc7
transformations.

The generic gauge field Lagrangian, parametrized as in \cite{dWST6},
was given in (\ref{eq:quadratic-L}) and contains moment couplings of
the field strength $F_{\mu\nu}{}^\Lambda$ with an operator
$\mathcal{O}_{\mu\nu\Lambda}$ which is quadratic in the fermions. Here
we will discuss the explict form of $\mathcal{O}_{\mu\nu\Lambda}$ and
of $\mathcal{N}_{\Lambda\Sigma}$. We start from the 56 field strengths
$G_{\mu\nu}{}^M$, introduced in susbsection \ref{sec:special-basis},
which transform under the \Exc7 transformations, which are embedded in
the $\mathrm{Sp}(56,\mathbb{R})$ electric/magnetic duality group. From
these field strengths and $\vv_M{}^{ij}$ and its complex conjugate, we
can construct \Exc7 invariant tensors. Specifically, 
consider the 56 \Exc7 invariant tensors, $\vv_M{}^{ij}
G^+_{\mu\nu}{}^M$ and $\vv_{M\,ij}G^+_{\mu\nu}{}^M$, and their
anti-selfdual counterparts that follow by hermitean conjugation. 
The fermionic bilinears $\mathcal{O}_{\mu\nu\Lambda}$ are
proportional to the following $\mathrm{SU}(8)$ covariant expression
\cite{deWitFreedman,deWit79,cremmer}, 
\begin{eqnarray}
  \label{eq:def-O}
  \mathcal{O}^+_{\mu\nu}{}^{ij} =   \ft12  \sqrt{2}\,
  \bar\psi_{\rho}^{i}\gamma^{[\rho}\gamma_{\mu\nu}\gamma^{\sigma]}
  \psi^{j}_{\sigma} 
  -\ft12\bar\psi_{\rho\,k}\gamma_{\mu\nu}\gamma^{\rho}\chi^{ijk}      
  - \ft1{144}\sqrt{2}\, \varepsilon^{ijklmnpq}\bar\chi_{klm}
  \gamma_{\mu\nu}\chi_{npq}  \,,
\end{eqnarray}
which is selfdual and transforming in the $\overline{\bf 28}$
representation of $\mathrm{SU}(8)$. Its complex conjugate is
anti-selfdual and transforms in the ${\bf 28}$ representation. The
fact that only a single tensor of fermionic bilinears appears in
the relation between the field strengths $F_{\mu\nu}{}^\Lambda$ and
the dual field strengths $G_{\mu\nu\Lambda}$, implies that this
relation must coincide with the following \Exc7 invariant 
equation,\footnote{ 
  We follow the argumentation presented in \cite{deWitNic}. The
  proportionality factor on the right-hand side of the equation
  follows from supersymmetry. } 
\begin{equation}
  \label{eq:field-strength-constraint}
  \vv_M{}^{ij}\, G^+_{\mu\nu}{}^M = -\ft1{2}
  \mathcal{O}^+_{\mu\nu}{}^{ij}\,. 
\end{equation}
The independent combination, $\vv_{M\,ij}G^+_{\mu\nu}{}^M$, defines an
$\mathrm{SU}(8)$ covariant tensor,
\begin{equation}
  \label{eq:def-F-ij}
  F^+_{\mu\nu\, ij}\equiv \vv_{M\,ij}\, G^+_{\mu\nu}{}^M\,,
\end{equation}
which will appear in the supersymmetry transformations of the
fermions. In this way both the \Exc7 invariance and the
$\mathrm{SU}(8)$ covariance of the supersymmetry transformations will
be ensured. Using (\ref{eq:VV-orthogonal}), we derive the following
equation,
\begin{equation}
  \label{eq:F-ij-G}
  G^+_{\mu\nu}{}^M=  \mathrm{i} \Omega^{MN}\Big[ \vv_{N}{}^{ij}
  F^+_{\mu\nu\,ij} +\ft12 \vv_{N\,ij} \mathcal{O}^+_{\mu\nu}{}^{ij} \Big]
  \,.  
\end{equation}
Furthermore, comparison of (\ref{eq:field-strength-constraint}) to
(\ref{eq:G-F-relation}) leads to a determination of
$\mathcal{N}_{\Lambda\Sigma}$ and $\mathcal{O}^+_{\mu\nu\Lambda}$,
\begin{eqnarray}
  \label{eq:N+O}
  \vv^{\Sigma ij}\,\mathcal{N}_{\Lambda\Sigma}&=& - 
  \vv_\Lambda{}^{ij}   \,,   \nonumber \\
  \vv^{\Lambda ij} \,\mathcal{O}^+_{\mu\nu\Lambda}&=& 
   \ft1{4}\mathrm{i} \,\mathcal{O}^+_{\mu\nu}{}^{ij} \,. 
\end{eqnarray}
These equations hold in any electric/magnetic duality frame and the
reader may verify that (\ref{eq:N-O-transform}) is indeed consistent
with (\ref{eq:V-em-duality}). Furthermore we note the relation,
\begin{equation}
  \label{eq:inv-Im-N}
  [(\mathcal{N} -\bar\mathcal{N})^{-1}]^{\Lambda\Sigma}
  = \mathrm{i}\,\vv^\Lambda{}_{ij} \,\vv^{\Sigma\,ij}\,.
\end{equation}
Observe that the imaginary part of the matrix $\mathcal{N}_{IJ}$ is
negative so that the kinetic term in (\ref{eq:quadratic-L}) carries
the correct sign. The sign in (\ref{eq:inv-Im-N}) depends crucially on
the sign adopted in (\ref{eq:VV-orthogonal}). We also note the
following relation 
\begin{equation}
  \label{eq:FO}
  F^+_{\mu\nu}{}^\Lambda\,\mathcal{O}^{+\mu\nu}_\Lambda = 
  - \ft14 F^+_{\mu\nu\,ij} \,\mathcal{O}^{+\mu\nu\,ij} 
  + 2\,
  \mathcal{O}^+_{\mu\nu\Lambda} \,\mathcal{O}^{+\mu\nu}_\Sigma \;
  \vv^\Lambda{}_{ij} \,\vv^{\Sigma\,ij} \,. 
\end{equation}

Most of the transformation rules and the Lagrangian can be deduced
from \cite{deWitNic}. As the reader may verify, they are consistent
with \Exc7 and $\mathrm{SU}(8)$ covariance. The transformation rules
can be written as follows,
\begin{eqnarray}
  \label{eq:susy-transformations}
  \delta \psi_{\mu}{}^{i}  &=&
  2\,\mathcal{D}_{\mu}\epsilon^{i}
  +\ft{1}{4}\sqrt{2}\,\hat{F}^{-}_{\rho\sigma}{}^{ij}\,
  \gamma^{\rho\sigma}\gamma_{\mu}\epsilon_{j}  +\ft14 \bar\chi^{ikl}
  \gamma^a\chi_{jkl} \; \gamma_a\gamma_\mu\epsilon^j 
  \nonumber\\
  &&
   + \ft12 \sqrt{2} \,\bar\psi_{\mu k}
  \gamma^a\chi^{ijk} \;\gamma_a\epsilon_j 
  - \ft1{576} \,\varepsilon^{ijklmnpq}
  \bar\chi_{klm}\gamma^{ab}\chi_{npq}
  \;\gamma_\mu\gamma_{ab}\epsilon_j 
  \;,   \nonumber\\[1ex]
  \delta \chi^{ijk} 
  &=&
  -2\sqrt{2}\,\hat{\cal P}_{\mu}^{ijkl}\,\gamma^{\mu}\epsilon_{l}
  + \ft3{2} \, \hat{F}^{-}_{\mu\nu}{}^{[ij}\gamma^{\mu\nu}\epsilon^{k]}
  - \ft1{24} \sqrt{2} \,\varepsilon^{ijklmnpq}
  \bar\chi_{lmn}\chi_{pqr} \; \epsilon^r  
  \;,   \nonumber\\[1ex]
  \delta e_{\mu}{}^{a}&=&
  \bar\epsilon^{i}\gamma^{a}\psi_{\mu i} ~+~
  \bar\epsilon_{i}\gamma^{a}\psi_{\mu}{}^i \;, \nonumber\\[1ex]
  \delta\vv_M{}^{ij} &=& 2\sqrt{2}\,\vv_{M kl} \, \Big(
  \bar\epsilon^{[i}\chi^{jkl]}+\ft1{24}\varepsilon^{ijklmnpq}\, 
  \bar\epsilon_{m}\chi_{npq}\Big)   \,,  \nonumber \\[1ex]
    \delta A_{\mu}{}^{M}
    &=&
    - \mathrm{i}\,\Omega^{MN} \vv_N{}^{ij}\,\Big( 
    \bar\epsilon^{k}\,\gamma_{\mu}\,\chi_{ijk}
    +2\sqrt{2}\, \bar\epsilon_{i}\,\psi_{\mu j}\Big)~+~ {\rm h.c.}
    \;. 
\end{eqnarray}
Here and henceforth the caret indicates that the corresponding
quantity is covariantized with respect to supersymmetry. For
completeness we record the expressions for
$\hat\mathcal{P}_\mu{}^{ijkl}$ and $\hat F^+_{\mu\nu ij}$ below,
\begin{eqnarray}
  \label{eq:supercov-P+F}
  \hat{\mathcal{P}}_\mu{}^{ijkl}&=&
  \mathcal{P}_\mu{}^{ijkl}-\sqrt{2}\,(\bar{\psi}^{[i}_\mu\,\chi^{jkl]}+
  \ft{1}{24}\,\varepsilon^{ijklmnpq}\,\bar{\psi}_{\mu m}\,\chi_{npq})
  \,,\nonumber\\   
  \hat{F}^+_{\mu\nu\,ij}&=& 
  F^+_{\mu\nu\,ij}+\ft{1}{4}\,\bar{\psi}_\rho{}^k
  \gamma^\rho\gamma_{\mu\nu}\chi_{ijk}-\ft{1}{8}\sqrt{2}\,
  \bar{\psi}_{\rho\,i}\{\gamma_{\mu\nu},\gamma^{\rho\sigma}\}
  \psi_{\sigma j}\,. 
\end{eqnarray}
The supercovariantized field strenghts $\hat G_{\mu\nu}{}^M$ then follow
from (\ref{eq:F-ij-G}) by substituting the second expression on the 
right-hand side, 
\begin{equation}
  \label{eq:scov-F-ij-G}
  \hat G^+_{\mu\nu}{}^M=  \mathrm{i} \Omega^{MN}\Big[ \vv_{N}{}^{ij}
  \hat F^+_{\mu\nu\,ij} -\ft1{288}\sqrt{2}\, \vv_{N\,ij} 
  \varepsilon^{ijklmnpq}\, \bar\chi_{klm} \gamma_{\mu\nu} \chi_{npq}
  \Big]    \,.  
\end{equation}

The derivatives $\mathcal{D}_\mu$ are covariant with respect to
Lorentz transformations and $\mathrm{SU}(8)$. For instance, we note,
\begin{equation}
  \label{eq:D-epsilon}
  \mathcal{D}_\mu \epsilon^i= \partial_\mu \epsilon^i
  -\ft1{4}\omega_\mu{}^{ab} \gamma_{ab}\epsilon^i + \ft1{2}
  \mathcal{Q}_\mu{}^i{}_j \epsilon^j\,. 
\end{equation}
The spin connection field $\omega_\mu{}^{ab}$ is consistent with the
expression one would obtain in first-order formalism, and corresponds
to the following value for the torsion tensor,
\begin{equation}
  \label{eq:spin-connection}
  \mathcal{D}_\mu e_\nu{}^a -\mathcal{D}_\nu e_\mu{}^a =\bar
  \psi_{[\mu}{}^i\gamma^a \psi_{\nu]i} +
  \ft1{12}\,\varepsilon_{\mu\nu}{}^{ab} \,\bar\chi^{ijk}\gamma_b
  \chi_{ijk} \,.  
\end{equation}
Note that we wrote down transformation rules for both electric and
magnetic gauge fields $A_\mu{}^M$. However, the (ungauged) Lagrangian
that we are about to introduce below does not depend on the magnetic
gauge fields $A_{\mu\Lambda}$. In view of what will happen when a
gauging is introduced, we will resolve this by assuming that the
Lagrangian is simply invariant under an additional local gauge
symmetry which acts exclusively on the magnetic gauge fields according
to $\delta A_{\mu\Lambda}= \Xi_{\mu\Lambda}$, where the
$\Xi_{\mu\Lambda}$ are independent space-time dependent functions.  At
this stage this may sound somewhat trivial, but the relevance of this
approach will become clear shortly when switching on general gaugings.

The above transformations (\ref{eq:susy-transformations}) close under
commutation. In particular the commutator of two consecutive
supersymmetry transformations $\delta(\epsilon_1)$ and
$\delta(\epsilon_2)$ leads to the following bosonic symmetry
variations,
\begin{equation}
  \label{eq:susy-algebra1}
  {}[\delta(\epsilon_1),\delta(\epsilon_2)] = \xi^\mu \hat D_\mu +
  \delta_{\rm L}(\epsilon^{ab}) + \delta_{\rm susy}(\epsilon_3) +
  \delta_{\rm SU(8)}(\Lambda^i{}_j) +  \delta_{\rm gauge}(\Lambda^M) +
  \delta_{\rm shift}(\Xi)  \,. 
\end{equation}
The first term indicates a general coordinate transformation, with
parameter $\xi^\mu$ given by
\begin{equation}
  \label{eq:gct-comutator}
  \xi^\mu = 2 (\bar \epsilon_2{}^i\gamma^\mu\epsilon_{1i} +\bar
  \epsilon_{2i} \gamma^\mu\epsilon_1{}^i)  \,,
\end{equation} 
whose covariantized form is generated on the matter fields by a
supercovariant derivative. The supersymmetry transformation parameter
$\epsilon_3$ is equal to 
\begin{equation}
  \label{eq:epsilon-3}
  \epsilon_{3 i} = -\sqrt{2}\, (\bar\epsilon_2{}^j
  \epsilon_1{}^k)\,\chi_{ijk} \,.
\end{equation}
The gauge transformation on the abelian gauge fields is expressed in
terms of the parameter, 
\begin{equation}
  \label{eq:gauge-commutator}
  \Lambda^M= - 4\mathrm{i} \sqrt{2}\, \Omega^{MN}\,(\vv_N{}^{ij}
  \,\bar\epsilon_{2i} \epsilon_{1j} - \vv_{Nij} \,\bar\epsilon_2{}^i
  \epsilon_1{}^j  )\,,
\end{equation}
which contributes to both electric and magnetic gauge fields. For
these fields the general coordinate transformation appears in the form
$- \xi^\nu \,G_{\mu\nu}{}^M$. For the electric gauge field the
$G_{\mu\nu}{}^\Lambda$ represents the standard field strength and this
term can be written as the linear combination of a general coordinate
transformation accompanied by a field-dependent gauge transformation.
For the magnetic gauge fields one can take the same point of view,
assuming that $G_{\mu\nu\Lambda}$ is actually the curl of
$A_{\mu\Lambda}$, which is a priori possible because the equations of
motion ({\it c.f.} (\ref{eq:eom-bianchi})) imply that
$G_{\mu\nu\Lambda}$ is subject to a Bianchi identity.  However, one
does not have to take this point of view, as the shift transformation
in (\ref{eq:susy-algebra1}), which acts exclusively on the magnetic
gauge fields, can always accomodate any terms that arise in the
supersymmetry commutator on $A_{\mu\Lambda}$.

We refrain from quoting any results for the parameters of the Lorentz
and the $\mathrm{SU}(8)$ transformations, as they will not play an
important role in what follows. In subsection
\ref{sec:gauged-supergravity} we return to the same supersymmetry
commutator in the presence of electric and magnetic charges and work
out some of the results in more detail.

The full Lagrangian for the ungauged theory can be written as follows,
\begin{eqnarray}
  \label{eq:ungauged-L}
   {\cal L} &=&
   -\ft12e\,R -\ft12\varepsilon^{\mu\nu\rho\sigma}\,
   (\bar\psi_{\mu}{}^{i}\gamma_{\nu}{D}_{\rho}\psi_{\sigma\,i}-  
   \bar\psi_{\mu}{}^{i}
   \overleftarrow{D}_{\rho}\gamma_{\nu}\psi_{\sigma\,i}) 
   \nonumber\\[1ex]
   &&{}
   -\ft14 \, \mathrm{i}e \Big\{\mathcal{N}_{\Lambda\Sigma} 
   \, {F}_{\mu\nu}^+{}^{\Lambda}\, {F}^{+\mu\nu\Sigma}  -
   \bar\mathcal{N}_{\Lambda\Sigma} \, {F}_{\mu\nu}^-{}^{\Lambda}\, 
   {F}^{-\mu\nu\Sigma} \Big\} \nonumber \\[1ex]
   &&{}
   -\ft1{12}e\,
   (\bar\chi^{ijk}\gamma^{\mu}D_{\mu}\chi_{ijk}-
   \bar\chi^{ijk}\overleftarrow{D}_{\mu}\gamma^{\mu}\chi_{ijk})
   -\ft1{12}e\,\vert {\cal P}_{\mu}^{ijkl}\vert^2 
   \nonumber\\[1ex]
   &&{}
   -\ft1{12} \sqrt{2}\,e\,\Big\{
   \bar\chi_{ijk}\gamma^{\nu}\gamma^{\mu}\psi_{\nu\,l}\,
   ({\cal P}_{\mu}^{ijkl}+ \hat{\cal P}_{\mu}^{ijkl}) + {\rm h.c.}
   \Big\}   \nonumber\\[1ex]
&&{} 
  +e {F}_{\mu\nu}^{+\Lambda}
  \,\mathcal{O}_\Lambda^{+\mu\nu} + e {F}_{\mu\nu}^{-\Lambda}
  \,\mathcal{O}_\Lambda^{-\mu\nu} 
  - e \vv^\Lambda{}_{ij}\,\vv^{\Sigma ij} \;\Big[
  \mathcal{O}_{\mu\nu\Lambda}^+ \mathcal{O}^{+\mu\nu}_{\Sigma}+ 
  \mathcal{O}_{\mu\nu\Lambda}^- \mathcal{O}^{-\mu\nu}_{\Sigma}\Big]
   \nonumber\\[1ex] 
   &&{}
   + \mathcal{L}_4 \;,  
\end{eqnarray}  
where $\mathcal{L}_4$ contains the following $\mathrm{SU}(8)$ invariant
terms quartic in the fermion fields, 
\begin{eqnarray}
  \label{eq:L-4}
  \mathcal{L}_4&=& 
  -\ft12e \bar \psi_\mu{}^{[i}\psi_\nu{}^{j]} \, \bar
  \psi^\mu{}_{i}\psi^\nu{}_{j} \,   \nonumber \\[1ex]
  &&{}
  + \ft18e \Big[\bar\psi_\rho{}^{k}\gamma^{\mu\nu}
  \gamma^\rho\chi_{ijk}\;  
  (\sqrt{2}\, \bar\psi_\mu{}^{i}\psi_\nu{}^{j} +\ft12 \bar 
  \psi_{\mu l}\gamma_\nu \chi^{ijl}) ~+~{\rm h.c.} \Big] 
  \nonumber\\[1ex]
  &&{}
  + \ft1{288} e \Big[ \varepsilon_{ijklmnpq} \bar \chi^{ijk}
  \gamma^{\mu\nu} 
  \chi^{lmn} \, ( \bar\psi_\mu{}^{p}\psi_\nu{}^{q} + \ft16\sqrt{2} \bar 
  \psi_{\mu r}\gamma_\nu \chi^{pqr})   ~+~{\rm h.c.} \Big]  
  \nonumber\\[1ex]
  &&{}
  +\ft1{32}e  \bar\chi^{ikl} \gamma^{\mu}\chi_{jkl}\, \bar\chi_{imn}
  \gamma_{\mu}\chi^{jmn} -\ft1{96} (\bar\chi^{ijk} \gamma^\mu
  \chi_{ijk})^2 \,. 
\end{eqnarray}
The terms of higher order in the fermions were taken from
\cite{deWitNic}, where their correctness was established in the
presence of the $\mathrm{SO}(8)$ gauging. However, in the
corresponding calculations only the generic properties of the
$T$-tensor were used, which do not depend on the choice of the gauge
group.  Hence these four-fermion terms must be universal. Observe that
the above Lagrangian applies to any electric/magnetic duality frame
because we can simply redefine the fields $\vv_M{}^{ij}$ by an 
$\mathrm{Sp}(56,\mathbb{R})$ matrix. 

\subsection{Introducing electric and magnetic charges}
\label{sec:electric-magnetic-charges}
Charges $X_M$ that couple to the gauge fields $A_\mu{}^M$ are
introduced in the standard way by extending covariant derivatives
according to (\ref{eq:cov-der}). In principle we include both electric
and magnetic charges and therefore we need both electric gauge fields
$A_\mu{}^\Lambda$ and magnetic gauge fields $A_{\mu\Lambda}$. The fact
that the latter did not appear so far in the Lagrangian
(\ref{eq:ungauged-L}) will not immediately pose a problem, but a
gauging usually induces a breaking of supersymmetry. Most of the
covariant derivatives do not lead to new terms when establishing
supersymmmetry, but there are variations involving the commutator of
the covariant derivatives that, in the presence of the gauging, lead
to the (nonabelian) field strengths $\mathcal{F}_{\mu\nu}{}^M$ defined
in (\ref{eq:field-strength}).  These terms, which are proportional to
the gauge coupling constant $g$, are easy to identify, as they
originate exclusively from the fermion kinetic terms. They are induced
the Cartan-Maurer equations (\ref{eq:GECM-Q-P}) whose right-hand sides
exhibit the extra terms proportional to the gauge coupling constant
$g$.  Collecting these terms leads to the following new variations,
\begin{eqnarray}
  \label{eq:first-var-L}
  && \delta\mathcal{L} = - e\, g\,\mathcal{H}_{\mu\nu}{}^M  \Big[
  \ft14 Q_{M i}{}^j (\bar\epsilon^i\gamma^\rho \gamma^{\mu\nu}
  \psi_{\rho j} - \bar\epsilon_j\gamma^{\mu\nu}\gamma^\rho
  \psi_{\rho}{}^i ) \nn\\
  &&{}\hspace{28mm}     + \ft1{144}\sqrt{2}\,
  \mathcal{P}_{M ijkl}\,\varepsilon^{ijklmnpq} \,
  \bar\chi_{mnp}\gamma^{\mu\nu}\epsilon_q \Big] + \mathrm{h.c.} \,, 
\end{eqnarray}
where $\mathcal{Q}_{M i}{}^j$ and $\mathcal{P}_{M ijkl}$ were defined
in (\ref{eq:P-Q-gauged}) and the replacement of
$\mathcal{F}_{\mu\nu}{}^M$ by $\mathcal{H}_{\mu\nu}{}^M$ is based on
(\ref{eq:Z-Q-P}).

It is, in principle, well known how these variations can be cancelled
\cite{deWitNic}.  Namely, one introduces masslike terms and new
supersymmetry variations for the fermions. These modifications
generate (among other terms) precisely the type of variations that may
cancel (\ref{eq:first-var-L}). The masslike terms are written as
follows,
\begin{eqnarray}
  \label{eq:masslike}
  \mathcal{L}_{\mathrm{masslike}} &=& 
   e\,g \,\Big\{\ft12\sqrt{2}\,  A_{1\,ij}\,
  \bar{\psi}_\mu{}^i 
  \gamma^{\mu\nu} \psi_\nu{}^j + \ft{1}{6}  A_{2i}{}^{jkl} \,
  \bar{\psi}_\mu{}^i \gamma^\mu \chi_{jkl}
  + A_3^{ijk,lmn}\,
  \bar{\chi}_{ijk}\,\chi_{lmn} \Big \}\nonumber \\
  &&{} 
  + ~{\rm h.c.} \,,
\end{eqnarray}
where
\begin{equation}
  \label{eq:A-3}
  A_3{}^{ijk,lmn} = \ft1{144} \sqrt{2}\,
  \varepsilon^{ijkpqr[lm}\,A_2{}^{n]}{}_{pqr} \,,
\end{equation}
and the new fermion variations are equal to 
\begin{eqnarray}
  \label{eq:delta-g-fermion}
  \delta_g \psi_{\mu}{}^{i}  &=&   
  {\sqrt{2}}\,g \,A_1{}^{ij}\,\gamma_{\mu}\,\epsilon_{j}
  \,,\nonumber\\ 
  \delta_g \chi^{ijk} &=& 
  - 2 g\,A_{2 l}{}^{ijk}\,\epsilon^{l}   \;. 
\end{eqnarray}
Here $A_1$ and $A_2$ are the components of the $T$-tensor defined in
(\ref{eq:su8-repre-constraint}). 

Furthermore we replace the abelian field strengths in the Lagrangian
(\ref{eq:ungauged-L}) by the field strengths
$\mathcal{H}_{\mu\nu}{}^\Lambda$, as described in
subsection \ref{sec:antisymmetric-tensors}, and we include the
topological and Chern-Simons-like terms (\ref{eq:topological}). In the
supersymmetry variations of the fermions we replace $F_{\mu\nu ij}$
accordingly by a tensor $\mathcal{H}_{\mu\nu ij}$ defined in analogy
with (\ref{eq:def-F-ij}), 
\begin{equation}
  \label{eq:def-H-ij}
  \mathcal{H}^+_{\mu\nu\, ij}\equiv \vv_{M\,ij}\,
  \mathcal{G}^+_{\mu\nu}{}^M\,. 
\end{equation}
Likewise we note three more relations, 
\begin{eqnarray}
  \label{eq:new-field-strength-constraint}
  \vv_M{}^{ij}\, \mathcal{G}^+_{\mu\nu}{}^M &=& -\ft1{2}
  \mathcal{O}^+_{\mu\nu}{}^{ij}\,, \\
  \label{eq:H-ij-G}
  \mathcal{G}^+_{\mu\nu}{}^M &=&  \mathrm{i} \Omega^{MN}\Big[
  \vv_{N}{}^{ij} 
  \mathcal{H}^+_{\mu\nu\,ij} +\ft12 \vv_{N\,ij}
  \mathcal{O}^+_{\mu\nu}{}^{ij} \Big]  \,,  \\
  \label{eq:new-FO}
  \mathcal{H}^+_{\mu\nu}{}^\Lambda\,\mathcal{O}^{+\mu\nu}_\Lambda &=& 
  - \ft14 \mathcal{H}^+_{\mu\nu\,ij} \,\mathcal{O}^{+\mu\nu\,ij} 
  + 2\,
  \mathcal{O}^+_{\mu\nu\Lambda} \,\mathcal{O}^{+\mu\nu}_\Sigma \;
  \vv^\Lambda{}_{ij} \,\vv^{\Sigma\,ij} \,, 
\end{eqnarray}
in direct analogy with (\ref{eq:field-strength-constraint}), 
(\ref{eq:F-ij-G}) and (\ref{eq:FO}), respectively. 

These above modifications generate a number of terms similar to
(\ref{eq:first-var-L}) originating from the fermion variations
proportional to $\mathcal{H}_{\mu\nu ij}$ in (\ref{eq:masslike}) and
from the fermion variations (\ref{eq:delta-g-fermion}) in the terms
$\mathcal{H}_{\mu\nu}{}^\Lambda \mathcal{O}^{\mu\nu}{}_\Lambda$ in the
original Lagrangian (\ref{eq:ungauged-L}) (upon the replacement of the
abelian field strengths by the $\mathcal{H}_{\mu\nu}{}^\Lambda$).
Dropping terms of higher order in the fermions, these variations take
the following form (here we also make use of
(\ref{eq:su8-repre-constraint})), 
\begin{eqnarray}
  \label{eq:second-var-L}
  &&\delta\mathcal{L} = e\, g\,\mathcal{H}^+_{\mu\nu \,kl}
  \Big[\ft13 T_i{}^{jkl}  (\bar\epsilon^i\gamma^\rho \gamma^{\mu\nu}
  \psi_{\rho j}-\bar\epsilon_j\gamma^{\mu\nu}\gamma^\rho
  \psi_{\rho}{}^i ) \nn\\
  &&{}\hspace{28mm}
  + \ft1{72}\sqrt{2}\,T_{mnpq}{}^{kl}\,  \varepsilon^{mnpqrstu} \,  
  \bar\chi_{rst}\gamma^{\mu\nu}\epsilon_u \Big]   + \mathrm{h.c.} \,. 
\end{eqnarray}
Using the definition of the $T$-tensor (\ref{eq:T1-tensor}) one can
show that (\ref{eq:second-var-L}) and (\ref{eq:first-var-L}) combine
to the expression,
\begin{eqnarray}
  \label{eq:third-var-L}
  && \delta\mathcal{L} = - e\, g\,\Big[\mathcal{H}^+_{\mu\nu}{}^M-
  \mathcal{G}^+_{\mu\nu}{}^M   \Big]\nn \\  
  && {} 
  \times \Big[\ft14 Q_{M i}{}^j (\bar\epsilon^i\gamma^\rho
  \gamma^{\mu\nu} \psi_{\rho j} - 
  \bar\epsilon_j\gamma^{\mu\nu}\gamma^\rho\psi_{\rho}{}^i )      
  + \ft1{6}\sqrt{2}\,
  \mathcal{P}_M{}^{ijmn}\, \bar\epsilon_i \gamma_{\mu\nu} \chi_{jmn} 
  \Big] + \mathrm{h.c.} \,,  
\end{eqnarray}
up to higher-order fermion terms. Here we made use of
(\ref{eq:H-ij-G}). For the electric components, where $M$ is replaced
by $\Lambda$, this term vanishes as one can read off from
(\ref{eq:cal-G}). The magnetic components can be cancelled by
assigning a suitable supersymmetry variation to the tensor fields.
Making use of (\ref{eq:B-field-eq}) one can determine this variation
directly,
\begin{eqnarray}
  \label{eq:Theta-delta-B}
    \Theta^{\Lambda \alpha} \,\delta B_{\mu\nu\,\alpha} &=&\mathrm{i}
    \Big(\ft{2}{3} \sqrt{2} \,\mathcal{P}^\Lambda{}_{ijkl} \,
    \bar\epsilon^{[i}\,\gamma_{\mu\nu}\,\chi^{jkl]}
    + 4\,\mathcal{Q}^\Lambda{}_{j}{}^i \,
    \bar\epsilon_{i}\,\gamma_{[\mu}\,\psi_{\nu]}{}^{j}
    ~-~ {\rm h.c.}\Big)  \nonumber\\
    &&{} 
    -2\,X^\Lambda{}_N{}^{P}\Omega_{PQ}\,A_{[\mu}{}^{N}\,\delta
    A_{\nu]}{}^{Q} \;.
\end{eqnarray}

At this point we have obtained a fairly complete version of all the
supersymmetry transformations. In principle one can now continue and
verify the cancellation of other variations of the Lagrangian. The
pattern of cancellations is very similar to the pattern exhibited in
\cite{deWitNic}. In the following subsection we first summarize the
full supersymmetry transformations and give the complete action. When
comparing the results to those for the electric gaugings, the
transformation rules for the magnetic gauge fields and the tensor
fields do not enter. To verify the completeness of these
transformation rules we will therefore verify the closure of the
supersymmetry commutator for all the bosonic fields. This commutator
will differ from (\ref{eq:susy-algebra1}), as there will be extra
terms related to the gauge transformations and furthermore the shift
transformation $\delta_{\mathrm{shift}}$ is replaced by the tensor
gauge transformations.

\subsection{Gauged maximal supergravity}
\label{sec:gauged-supergravity}
In this section we present the complete results for gauged
supergravity. The supersymmetry transformation rules turn out to take
the following form,
\begin{eqnarray}
  \label{eq:susy-transformations-gauged}
  \delta \psi_{\mu}{}^{i} 
  &=&
  2\,\mathcal{D}_{\mu}\epsilon^{i}
  +\ft{1}{4}\sqrt{2}\,\hat{\mathcal{H}}^{-}_{\rho\sigma}{}^{ij}\,
  \gamma^{\rho\sigma}\gamma_{\mu}\epsilon_{j}  +\ft14 \bar\chi^{ikl}
  \gamma^a\chi_{jkl} \; \gamma_a\gamma_\mu\epsilon^j 
  \nonumber\\
  &&
   + \ft12 \sqrt{2} \,\bar\psi_{\mu k}
  \gamma^a\chi^{ijk} \;\gamma_a\epsilon_j 
  - \ft1{576} \varepsilon^{ijklmnpq}
  \bar\chi_{klm}\gamma^{ab}\chi_{npq}
  \;\gamma_\mu\gamma_{ab}\epsilon_j  \nonumber\\
  && +{\sqrt{2}}\,g \,A_1{}^{ij}\,\gamma_{\mu}\,\epsilon_{j} \;,
     \nonumber\\[1ex]
  \delta \chi^{ijk} 
  &=&
  -2\sqrt{2}\,\hat{\cal P}_{\mu}^{ijkl}\,\gamma^{\mu}\epsilon_{l}
  + \ft3{2} \, \hat{\mathcal{H}}^{-}_{\mu\nu}{}^{[ij}
  \gamma^{\mu\nu}\epsilon^{k]} 
  - \ft1{24} \sqrt{2} \,\varepsilon^{ijklmnpq}
  \bar\chi_{lmn}\chi_{pqr} \; \epsilon^r  \nonumber\\
  &&{}
  - 2 g\,A_{2 l}{}^{ijk}\,\epsilon^{l}   \;,   \nonumber\\[1ex]
  \delta e_{\mu}{}^{a}&=&
  \bar\epsilon^{i}\gamma^{a}\psi_{\mu i} ~+~
  \bar\epsilon_{i}\gamma^{a}\psi_{\mu}{}^i \;, \nonumber\\[1ex]
  \delta\vv_M{}^{ij} &=&  2\sqrt{2}\,\vv_{M kl} \, \Big(
  \bar\epsilon^{[i}\chi^{jkl]}+\ft1{24}\varepsilon^{ijklmnpq}\, 
  \bar\epsilon_{m}\chi_{npq}\Big)   \,,  \nonumber \\[1ex]
    \delta A_{\mu}{}^{M}
    &=&
    - \mathrm{i}\,\Omega^{MN} \vv_N{}^{ij}\,\Big( 
    \bar\epsilon^{k}\,\gamma_{\mu}\,\chi_{ijk}
    +2\sqrt{2}\, \bar\epsilon_{i}\,\psi_{\mu j}\Big)~+~ {\rm h.c.}
    \;, \nonumber \\[1ex]
    \delta B_{\mu\nu\,\alpha} &=&\ft{2}{3} \sqrt{2} \,
    t_{\alpha M}{\!}^{P}\Omega^{MQ}\, \Big( {\cal V}_{P\,ij} {\cal
    V}_{Q\,kl}\, 
    \bar\epsilon^{[i}\,\gamma_{\mu\nu}\,\chi^{jkl]}
    + 2 \sqrt{2}\, {\cal V}_{P\,jk} {\cal V}_{Q}{}^{ik}\,
    \bar\epsilon_{i}\,\gamma_{[\mu}\,\psi_{\nu]}{}^{j}
    ~+~ {\rm h.c.}\Big)  \nonumber\\
    &&{} 
    -2t_{\alpha\,M}{}^{P}\Omega_{PN}\,A_{[\mu}{}^{M}\,\delta
    A_{\nu]}{}^{N} \;.
\end{eqnarray}
As was already noted before (see the text preceding
(\ref{eq:B-transf-0})), we only need the variations $\Theta_M{}^\alpha
\,\delta B_{\mu\nu\alpha}$, which can conveniently be written as,
\begin{eqnarray}
  \label{eq:Theta-delta-B-proj}
    \Theta_M{}^\alpha \,\delta B_{\mu\nu\,\alpha} &=&\mathrm{i}
    \Big(\ft{2}{3} \sqrt{2} \,\mathcal{P}_{Mijkl} \,
    \bar\epsilon^{[i}\,\gamma_{\mu\nu}\,\chi^{jkl]}
    + 4\,\mathcal{Q}_{Mj}{}^i \,
    \bar\epsilon_{i}\,\gamma_{[\mu}\,\psi_{\nu]}{}^{j}
    ~-~ {\rm h.c.}\Big)  \nonumber\\
    &&{} 
    -2\,X_{MN}{}^{P}\Omega_{PQ}\,A_{[\mu}{}^{N}\,\delta
    A_{\nu]}{}^{Q} \;.
\end{eqnarray}
The above variations were determined by the substitution of
$\mathcal{H}^+_{\mu\nu ij}$ for $F^+_{\mu\nu ij}$ into
(\ref{eq:susy-transformations}) and by including the variations
(\ref{eq:delta-g-fermion}). For the tensor field $B_{\mu\nu\alpha}$ we
based ourselves on (\ref{eq:Theta-delta-B}).

At this point we return to the commutator of two supersymmetry
transformations, which still takes the form (\ref{eq:susy-algebra1}), but now
with the last `shift' transformation on the magnetic gauge fields
replaced by a full tensor gauge transformation,
\begin{equation}
  \label{eq:susy-algebra2}
  {}[\delta(\epsilon_1),\delta(\epsilon_2)] = \xi^\mu \hat D_\mu +
  \delta_{\rm L}(\epsilon^{ab}) + \delta_{\rm susy}(\epsilon_3) +
  \delta_{\rm SU(8)}(\Lambda^i{}_j) +  \delta_{\rm gauge}(\Lambda^M) +
  \delta_{\rm tensor}(\Xi_{\mu\alpha})  \,. 
\end{equation}
As before, the first term represents a covariantized general
coordinate transformation, where one must now also include terms of
order $g$ induced by the gauging. The parameters $\epsilon_3$ and
$\Lambda^M$ of the supersymmetry and gauge transformations appearing
on the right-hand side, were already given in (\ref{eq:epsilon-3}) and
(\ref{eq:gauge-commutator}), respectively.

Because the magnetic vector and the tensor gauge fields are new as
compared to previous treatments, we briefly consider the realization of
(\ref{eq:susy-algebra2}) on the vector and tensor gauge fields. As a
non-trivial consistency check on our reuslt, we include all
higher-order fermion contributions in the supersymmetry commutator
acting on the vector fields. For the tensor gauge field we include all
bilinears in the fields $\chi^{ijk}$. In this way we also determine
the parameter of the tensor gauge transformation in
(\ref{eq:susy-algebra2}). On the vector gauge fields we derive,
\begin{eqnarray}
  \label{eq:gct-gauge-fields2}
  {}[\delta(\epsilon_1),\delta(\epsilon_2)] A_\mu{}^M &=& -\xi^\nu
  \left[\hat{\mathcal{G}}_{\mu\nu}{}^M  -\ft12 \mathrm{i}\,
  \Omega^{MN} \vv_{N}{}^{ij} \bar\psi_{\mu}{}^k \gamma_\nu\chi_{ijk}
  +\ft12 \mathrm{i}\, \Omega^{MN} \vv_{Nij}\bar\psi_{\mu k} 
  \gamma_\nu\chi^{ijk}   \right]
   \nonumber\\
  &&{} 
  + D_\mu \Lambda^M - gZ^{M,\alpha}\, \Xi_{\mu\alpha} +
  \delta(\epsilon_3)  \,, 
\end{eqnarray}
 where  
\begin{equation}
  \label{eq:Xi}
  \Theta_M{}^\alpha \,\Xi_{\mu\alpha} = - 4 \mathrm{i}
  \,\mathcal{Q}_{Mi}{}^j\,(\bar\epsilon_2{}^i \gamma_\mu \epsilon_{1j}
  + \bar\epsilon_{2j} \gamma_\mu \epsilon_1{}^i) \;,
\end{equation}
defines a contribution to the parameters of the tensor gauge
transformations.  These are not the only terms, as we will see by
evaluating the first term on the right-hand side of
(\ref{eq:gct-gauge-fields2}). We remind the reader that we are only
interested in the algebra acting on the fields $A_\mu{}^\Lambda$ and
$\Theta^{\Lambda\alpha}A_{\mu\Lambda}$, as was explained at the end of
subsection \ref{sec:antisymmetric-tensors}. This enables us to replace
$\Theta^{\Lambda\alpha} \hat{\mathcal{G}}_{\mu\nu\Lambda}$ by
$\Theta^{\Lambda\alpha} \hat{\mathcal{H}}_{\mu\nu\Lambda}$, by making
use of the field equations (\ref{eq:B-field-eq}) of the tensor field.
This result applies also to the supercovariant extensions of the field
strengths (this can be deduced from the observation that field
equations transform into field equations under a symmetry of the
action). Hence we must evaluate the expression,
\begin{eqnarray}
  \label{eq:xi-cal-H}
  -\xi^\nu\, \hat{\mathcal{H}}_{\mu\nu}{}^M &=& \xi^\nu\partial_\nu
   A_\mu{}^M + \partial_\mu \xi^\nu A_\nu{}^M  - D_\mu(\xi^\nu
   A_\nu{}^M)\nn\\ 
   &&{} 
   - \mathrm{i}\xi^\nu\,\Omega^{MN} \left[\vv_{N}{}^{ij}[ 
   \bar\psi_{[\mu}{}^k \gamma_{\nu]} \chi_{ijk} +\sqrt{2}
   \bar\psi_{\mu i}\psi_{\nu j}]  - \mathrm{h.c.}   \right] \nonumber\\
   &&{}
   - gZ^{M,\alpha} \,\xi^\nu\left[B_{\mu\nu\alpha} -
   t_{\alpha N}{}^Q\,\Omega_{PQ}\, A_\mu{}^N  A_\nu{}^P\right] 
   \,.
\end{eqnarray}
Combining this expression with the fermionic bilinears in
(\ref{eq:gct-gauge-fields2}) shows that the result decomposes into a
space-time diffeomorphism with parameter $\xi^\mu$, a nonabelian gauge
transformation with parameter $- \xi^\mu A_\mu{}^M$, a supersymmetry
transformation with parameter $-\ft12\xi^\nu \psi_{\nu i}$, and a
tensor gauge transformation with parameter $\xi^\nu( B_{\mu\nu\alpha}
- t_{\alpha N}{}^Q\,\Omega_{PQ}\, A_\mu{}^N A_\nu{}^P)$.

Subsequently we evaluate the supersymmetry commutator on the tensor
fields $\Theta_M{}^\alpha B_{\mu\nu\alpha}$. Including all terms
quadratic in $\chi^{ijk}$, we derive the following result,
\begin{eqnarray}
  \label{eq:B-commutator}
  {[}\delta(\epsilon_1),\delta(\epsilon_2)]\,\Theta_M{}^\alpha
  B_{\mu\nu\alpha}&=&   2\,\Theta_M{}^\alpha\, D_{[\mu} 
  \Xi_{\nu]\alpha}  +  2 \,X_{MN}{}^P \hat{\mathcal{G}}_{\mu\nu}{}^N
  \Omega_{PQ} \,\Lambda^Q  \nonumber\\ 
  &&{}  
    +\ft{2}{3}\mathrm{i} \sqrt{2} \left(\mathcal{P}_{Mijkl} \,
    \bar\epsilon_3^{[i}\,\gamma_{\mu\nu}\,\chi^{jkl]} - \mathrm{h.c.}
    \right) \nonumber \\
    &&{}
   +  \mathrm{i} e\,\varepsilon_{\mu\nu\rho\sigma}\,\xi^\sigma \left[ 
     \ft13\mathcal{P}_{M\,ijkl} \, \hat\mathcal{P}^{\rho\,ijkl}
     +\ft12 \mathcal{Q}_{Mi}{}^j\,\bar\chi^{ikl}\gamma^\rho\chi_{jkl}
  \right] 
  \nonumber\\
  &&{}
  -2\,X_{MN}{}^P\Omega_{PQ} \,A_{[\mu}{}^N
   \,{[}\delta(\epsilon_1),\delta(\epsilon_2)] A_{\nu]}{}^Q ~+~\cdots \,,
\end{eqnarray}
where $\xi^\mu$, $\Lambda^M$, $\epsilon_3$ and $\Xi_{\mu\alpha}$ have
already been given in (\ref{eq:gct-comutator}),
(\ref{eq:gauge-commutator}), (\ref{eq:epsilon-3}) and (\ref{eq:Xi}),
respectively, and the dots represent additional terms linear and
quadratic in $\psi_\mu{}^i$.  To derive this expression we used many
of the results obtained previously. We draw attention to the fact that
we also need the torsion constraint (\ref{eq:spin-connection}).
Obviously the commutator closes with respect to these parameters in
view of the fact that closure was already established on the gauge
fields $A_\mu{}^M$. Note also the second term proportional to
$\Lambda^Q$, which is implied by the last term shown in
(\ref{eq:B-transf-0}).

What remains is to investigate the closure relation for the terms
proportional to the parameter $\xi^\mu$ of the general coordinate
transformations. For these terms it is important to restrict ourselves
to the commutator on $\Theta^{\Lambda\alpha}B_{\mu\nu\alpha}$, as
these are the only components of the tensor field on which the
supersymmetry algebra should be realized (we refer to the discussion
at the end of subsection \ref{sec:antisymmetric-tensors}). We will
first show that closure is indeed achieved provided the following
equation holds,
\begin{equation}
  \label{eq:field-eq}
   \ft16\mathrm{i} g \,\varepsilon^{\mu\nu\rho\sigma}
   \mathcal{H}_{\nu\rho\sigma\alpha} \, \Theta^{\Lambda\alpha}
   + e g\left( \ft13\mathcal{P}^\Lambda{}_{ijkl} \,
     \hat\mathcal{P}^{\mu\,ijkl} +\ft12
     \mathcal{Q}^\Lambda{}_{i}{}^j\,\bar\chi^{ikl}\gamma^\mu\chi_{jkl}  
   \right) +\cdots ~=~ 0\,. 
\end{equation}
Here the unspecified terms are proportional to gravitino fields, which
are suppressed throughout this calculation.  This equation is simply
the field equation associated with the magnetic vector fields (up to
terms that vanish upon using the field equation for the tensor
fields). The first term was aready evaluated in
(\ref{eq:A-field-eq-2}) and the second term originates from the
minimal coulings which enter through $\mathcal{P}_M{}^{ijkl}$ and
$\mathcal{Q}_{\mu i}{}^j$. It is perhaps unexpected that the
supersymmetry algebra closes modulo a bosonic field equation that
involves space-time derivatives, but one has to bear in mind that
these particular field equations are only of first order in
derivatives. Using the above equation we derive,
\begin{eqnarray}
  \label{eq:P+Q-xi}
  \lefteqn{
  \mathrm{i} e\,\varepsilon_{\mu\nu\rho\sigma}\,\xi^\sigma \left[ 
     \ft13\mathcal{P}^\Lambda{}_{ijkl} \, \hat\mathcal{P}^{\rho\,ijkl}
     +\ft12 \mathcal{Q}^\Lambda{}_{i}{}^j\,\bar\chi^{ikl}\gamma^\rho\chi_{jkl}
  \right] =}
    \nonumber\\[.5ex]
  &&\qquad
  \Theta^{\Lambda\alpha}\left[ \xi^\rho\partial_\rho B_{\mu\nu\alpha}  -
  2\,\partial_{[\mu} \xi^\rho \, B_{\nu]\rho\alpha} \right] \nonumber\\
  &&\qquad
  +2 \,\Theta^{\Lambda\alpha} D_{[\mu} \left[ \xi^\rho (B_{\nu]\rho\alpha}
  - t_{\alpha N}{}^Q\,\Omega_{PQ}\, A_{\nu]}{}^N A_\rho{}^P) \right]
  \nonumber\\
  &&\qquad
  -2\,X^\Lambda{}_{N}{}^P \, \mathcal{G}_{\mu\nu}{}^N \, \Omega_{PQ}\, \xi^\rho
  A_\rho{}^Q \nonumber\\
  &&\qquad
  +2\,X^\Lambda{}_{N}{}^P \Omega_{PQ}
  \,A_{[\mu}{}^N\left[\xi^\rho\partial_\rho  A_{\nu]}{}^Q +
  \partial_{\nu]}\xi^\rho\,A_\rho{}^Q - 2\,\xi^\rho
  (\mathcal{G}-\mathcal{H})_{\nu]\rho}{}^Q \right]   \,.  
\end{eqnarray}
This establishes that full closure is indeed realized. The first line
represents the required general coordinate transformation, the second
and third term corresponds to the extra vector and tensor gauge
transformations, respectively, with the same parameters as found in
(\ref{eq:xi-cal-H}). Finally, the last term cancels against the
similar terms generated on $A_\nu{}^Q$ by the commutator in the last
term of (\ref{eq:B-commutator}). Here it is important to realize that
this commutator does not fully close, in view of the fact that
$A_\mu{}^Q$ includes all the magnetic gauge fields, as there is no
contraction with $\Theta_Q{}^{\alpha}$. Nevertheless one is still left
with a term proportional to $(\mathcal{G}-\mathcal{H})_{\nu\rho}{}^Q$,
which can be absorbed into a transformation of type
(\ref{eq:Delta-invariance}).

The full universal Lagrangian of maximal gauged supergravity in four
space-time dimensions reads as follows,
\begin{eqnarray}
  \label{eq:gauged-L}
   {\cal L} &=&
   -\ft12e\,R -\ft12\varepsilon^{\mu\nu\rho\sigma}\,
   (\bar\psi_{\mu}{}^{i}\gamma_{\nu}{D}_{\rho}\psi_{\sigma\,i}-  
   \bar\psi_{\mu}{}^{i}
   \overleftarrow{D}_{\rho}\gamma_{\nu}\psi_{\sigma\,i}) 
   \nonumber\\[1ex]
   &&
   -\ft14 \, \mathrm{i}e \Big\{\mathcal{N}_{\Lambda\Sigma} 
   \, \mathcal{H}_{\mu\nu}^+{}^{\Lambda}\, \mathcal{H}^{+\mu\nu\Sigma}  -
   \bar\mathcal{N}_{\Lambda\Sigma} \, \mathcal{H}_{\mu\nu}^-{}^{\Lambda}\, 
   \mathcal{H}^{-\mu\nu\Sigma} \Big\} \nonumber \\[1ex]
   &&{}
   -\ft1{12}e\,
   (\bar\chi^{ijk}\gamma^{\mu}D_{\mu}\chi_{ijk}-
   \bar\chi^{ijk}\overleftarrow{D}_{\mu}\gamma^{\mu}\chi_{ijk})
   -\ft1{12}e\,\vert {\cal P}_{\mu}^{ijkl}\vert^2 
   \nonumber\\[1ex]
   &&{}
   -\ft1{12} \sqrt{2}\,e\,\Big\{
   \bar\chi_{ijk}\gamma^{\nu}\gamma^{\mu}\psi_{\nu\,l}\,
   ({\cal P}_{\mu}^{ijkl}+ \hat{\cal P}_{\mu}^{ijkl}) + {\rm h.c.}
   \Big\}   \nonumber\\[1ex]
&&{} 
  +e \mathcal{H}_{\mu\nu}^{+\Lambda}
  \,\mathcal{O}_\Lambda^{+\mu\nu} + e \mathcal{H}_{\mu\nu}^{-\Lambda}
  \,\mathcal{O}_\Lambda^{-\mu\nu} 
  - e \vv^\Lambda{}_{ij}\,\vv^{\Sigma ij} \;\Big[
  \mathcal{O}_{\mu\nu\Lambda}^+ \mathcal{O}^{+\mu\nu}_{\Sigma}+ 
  \mathcal{O}_{\mu\nu\Lambda}^- \mathcal{O}^{-\mu\nu}_{\Sigma}\Big]
  \nonumber\\[1ex]
  &&{}
+\ft1{8}\mathrm{i} g\, \varepsilon^{\mu\nu\rho\sigma}\,
\Theta^{\Lambda\alpha}\,B_{\mu\nu\,\alpha} \,
\Big(
2\partial_{\rho} A_{\sigma\,\Lambda} + g
X_{MN\,\Lambda} \,A_\rho{}^M A_\sigma{}^N
-\ft14g\Theta_{\Lambda}{}^{\beta}B_{\rho\sigma\,\beta} \Big)
\nonumber\\[.9ex]
&&{}
+\ft1{3}\mathrm{i} g\, \varepsilon^{\mu\nu\rho\sigma}X_{MN\,\Lambda}\,
A_{\mu}{}^{M} A_{\nu}{}^{N}
\Big(\partial_{\rho} A_{\sigma}{}^{\Lambda}
+\ft14 gX_{PQ}{}^{\Lambda} A_{\rho}{}^{P}A_{\sigma}{}^{Q}\Big)
\nonumber\\[.9ex]
&&{}
+\ft1{6}\mathrm{i} g\, \varepsilon^{\mu\nu\rho\sigma}X_{MN}{}^{\Lambda}\,
A_{\mu}{}^{M} A_{\nu}{}^{N}
\Big(\partial_{\rho} A_{\sigma}{}_{\Lambda}
+\ft14 gX_{PQ\Lambda} A_{\rho}{}^{P}A_{\sigma}{}^{Q}\Big)
\nonumber\\[.9ex]
&&{}
+g\, e\Big\{\ft12\sqrt{2}\,  A_{1\,ij}\,
\bar{\psi}\,^i_\mu
\gamma^{\mu\nu} \psi^j_\nu + \ft{1}{6}  A_{2i}{}^{jkl} \,
\bar{\psi}^i_\mu \gamma^\mu \chi_{jkl}
+ A_3^{ijk,lmn}\,
\bar{\chi}_{ijk}\,\chi_{lmn} + {\rm h.c.}\Big \}
\nonumber\\[.9ex]
&&{}
-g^2 e\,\Big \{\ft{1}{24}  A_{2i}{}^{\!jkl}A^{\,\,i}_{2\,jkl} -
\ft{3}{4}  A_1^{ij}A^{\vphantom{j}}_{1\,ij}\Big\}
   \nonumber\\[1ex] 
   &&{}
   + \mathcal{L}_4 \;,  
\end{eqnarray}  
where $\mathcal{L}_4$ was given in (\ref{eq:L-4}). Here we included
the scalar potential which appears at order $g^2$ and which takes the
form already derived in \cite{deWitNic}. We note that this potential
can be written in various ways, 
\begin{eqnarray}
  \label{eq:g-potential}
  \mathcal{P}(\vv )&=& g^2 \Big \{\ft{1}{24}  \vert A_{2i}{}^{\!jkl}\vert^2 -
  \ft{3}{4} \vert A_1^{ij}\vert^2\Big\} \nonumber\\[1ex]
  &=& \ft{1}{336}\, g^2 \,\mathcal{M}^{MN} \left\{ 8\,
  \mathcal{P}_M{}^{ijkl} \mathcal{P}_{N ijkl} 
  + 9\, \mathcal{Q}_{Mi}{}^j\, \mathcal{Q}_{Nj}{}^i  \right\}  
  \nonumber\\[1ex]  
  &=& \ft1{672}\, g^2 \Big \{X_{MN}{}^{R}X_{PQ}{}^{S}{\cal M}^{MP}{\cal
  M}^{NQ}{\cal M}_{RS} + 7\,X_{MN}{}^{Q}X_{PQ}{}^{N}{\cal M}^{MP}
  \Big\} \,,
\end{eqnarray}
where we have used the real symmetric field-dependent $56\times56$
matrix ${\cal M}_{MN}$, defined by
\begin{equation}
  \label{eq:def-M}
  {\cal M}_{MN} \equiv
  \vv_M{}^{ij} \,\vv_{N\,ij} + \vv_{M\,ij}\, \vv_N{}^{ij}  \;.
\end{equation}
Note that $\mathcal{M}$ is positive
definite. Its inverse, $\mathcal{M}^{MN}$, can be written as
\begin{equation}
  \label{eq:Omega-M}
  {\cal M}^{MN}= \Omega^{MP} \Omega^{NQ}{\cal M}_{PQ} \,,
\end{equation}
by virtue of (\ref{eq:VV-orthogonal}). This shows that
$\det[\mathcal{M}] =1$. 

In the derivation of (\ref{eq:g-potential}) we made use of the
following equations, 
\begin{eqnarray}
  \label{eq:XX-PP-QQ}
  X_{MN}{}^{R}X_{PQ}{}^{S}{\cal M}^{MP}{\cal M}^{NQ}{\cal M}_{RS}&=&
  \mathcal{M}^{MN} \left(2\,\mathcal{P}_M{}^{ijkl} \mathcal{P}_{N ijkl}
  - 3\,\mathcal{Q}_{Mi}{}^j\, \mathcal{Q}_{Nj}{}^i  \right) \,,
  \nonumber\\ 
  X_{MN}{}^{Q}X_{PQ}{}^{N}{\cal M}^{MP} &=& 
    \mathcal{M}^{MN} \left(2\, \mathcal{P}_M{}^{ijkl} \mathcal{P}_{N ijkl}
  + 3\,\mathcal{Q}_{Mi}{}^j\, \mathcal{Q}_{Nj}{}^i  \right) \,,
  \nonumber \\
  \mathcal{M}^{MN} \mathcal{P}_M{}^{ijkl} \mathcal{P}_{N ijkl} &=& 
  4\, \vert A_{2l}{}^{ijk}\vert^2  \,,
  \nonumber\\ 
  \mathcal{M}^{MN} \mathcal{Q}_{Mi}{}^j\, \mathcal{Q}_{Nj}{}^i &=& 
  -2 \, \vert A_{2l}{}^{ijk}\vert^2 -28\, \vert A_1{}^{ij}\vert^2 \,,
\end{eqnarray}
which can be derived using various results and definitions presented
in section \ref{sec:T-tensor}.

\section{Discussion and applications}
\label{sec:disc-appl}
\setcounter{equation}{0}
In this paper we have presented the complete construction of all
gaugings of four-dimensional maximal supergravity. We have shown that
gaugings can be completely characterized in terms of an embedding
tensor, subject to a linear and a quadratic constraint,
(\ref{eq:repres-constraint}) and (\ref{closure-constraint}),
respectively.  A generic gauging can involve both electric and
magnetic charges, together with two-form tensor fields transforming in
the ${\bf 133}$ representation of \Exc7.  The addition of magnetic
vector fields and the two-rank tensor fields does not lead to
additional degrees of freedom owing to the presence of extra gauge
invariances associated with these fields.  We have presented the full
Lagrangian of the theory in (\ref{eq:gauged-L}) and the supersymmetry
transformations in (\ref{eq:susy-transformations-gauged}).

In this last section we briefly demonstrate the group-theoretical
approach of this paper to construct actual gaugings of maximal
supergravity in four dimensions.  The starting point is the
construction of a solution to the constraints
(\ref{eq:repres-constraint}) and (\ref{closure-constraint}) on the
embedding tensor $\Theta_M{}^\alpha$. The former one is a linear
constraint whose general solution is explicitly known as the
912-dimensional image of a projector.  The most straightforward
strategy will thus be to start from a particular solution to this
constraint and impose on it the quadratic constraint.

Of course, when one wants to see if a specific subgroup of \Exc7 can
be gauged, it suffices to simply verify whether the constraints are
satisfied on the corresponding embedding tensor. In other cases, when
one wants to explore a variety of gaugings, it is often useful to
first select a subgroup $\mathrm{G}_0\subset\mathrm{E}_{7(7)}$ in
which the gauge group will be embedded eventually. This group may be a
manifest invariance of the ungauged Lagrangian in a suitable
electric/magnetic duality frame. When this is the case, the gauging
will only involve electric gauge fields and there is no need for
introducing dual vector and tensor fields.  Branching the ${\bf 912}$
of ${\rm E}_{7(7)}$ under ${\rm G}_0$ and scanning through the
different irreducible components allows a systematic study of the
quadratic constraint~(\ref{closure-constraint}) and thereby a full
determination of the corresponding admissible gaugings. In that case
the closure of the gauge group is already guaranteed, owing to the
equivalent formulation ~(\ref{eq:closure-constraint2}) of the
quadratic constraint, so that every solution to the linear constraint
(\ref{eq:repres-constraint}) will define a viable gauging.

A central result of this paper is that it is not necessary to restrict
$\mathrm{G}_0$ to a group that can be realized as an invariance of the
ungauged Lagrangian that serves as a starting point for the gauging.
In that case, one must simply analyze both constraints and the gauging
may eventualy comprise both electric and magnetic charges. It is
important to realize that the scalar potential is insensitive to the
issue of electric/magnetic frames, so that its stationary point can be
directly studied.  Scanning through the different choices of ${\rm
  G}_0$, it is straightforward to construct the various corresponding
gaugings of the four-dimensional theory. In the following we will
illustrate the strategy by first reproducing the known gaugings,
subsequently sketching the construction of gaugings related to flux
compactifications of IIA and IIB supergravity and finally giving some
other examples, including Scherk-Schwarz reductions from higher
dimensions.
\subsection{Known gaugings}
As first examples, let us briefly review the known gaugings embedded
in the groups ${\rm G}_{\rm 0}={\rm SL}(8,\mathbb{R})$ and ${\rm
  G}_{\rm 0}={\rm E}_{6(6)}\times {\rm SO}(1,1)$, respectively. It is
known that there are corresponding ungauged Lagrangians which have
these groups as an invariance group. Hence we can restrict ourselves
to analyzing the linear constriant. With the group ${\rm G}_{\rm
  0}={\rm SL}(8,\mathbb{R})$, the branching of the ${\rm E}_{7(7)}$
representations associated with the vector fields, the adjoint
representation and the embedding tensor, is as follows, 
\begin{eqnarray}
{\bf 56}&\rightarrow & {\bf 28}+{\bf 28}'\,, \nonumber\\
{\bf 133}&\rightarrow & {\bf 63}+{\bf 70}\,, \nonumber\\
{\bf 912}&\rightarrow & {\bf 36}+{\bf 420}+ {\bf 36}'+ {\bf 420}' \,,
\label{sl8decs}
\end{eqnarray}
where the ${\bf 28}$ representation in the first decomposition denotes
the electric gauge fields. The possible couplings between vector
fields and ${\rm E}_{7(7)}$ symmetry generators induced by the various
$\Theta$ components according to~(\ref{eq:X-theta-t}) can be
summarized in the table
\begin{equation}
\begin{tabular}{c|cc}
&${\bf 28}$ &${\bf 28}'$
\\
\hline
${\bf 63}$
&${\bf 36}+{\bf 420}$
&${\bf 36}'+{\bf 420}'$
\\
${\bf 70}$
&${\bf 420}'$
&${\bf 420}$
\\
\end{tabular}
\label{decblock}
\end{equation}
where the left column represents the ${\rm E}_{7(7)}$ generators, and
the top row represents the vector fields. The entries correspond to
the {\em conjugate} representations of the respective components of
the embedding tensor belonging to the ${\bf 912}$ representation. 
Restricting to gaugings embedded into
$\mathrm{G}_0=\mathrm{SL}(8,\mathbb{R})$, the upper left entry is
relevant. However, the ${\bf 420}$ would alsocouple to the magnetic
gauge fields and the remaining generators of \Exc7 so that the embedding
tensor is restricted to live in the ${\bf 36}'$ (i.e. the {\it
conjugate} of the ${\bf 36}$ indicated in the table). Every element
in the ${\bf 36}'$  defines a viable 
gauging. A closer analysis shows \cite{dWST1} that modulo ${\rm
  SL}(8,\mathbb{R})$ conjugation the general form of $\Theta\in{\bf
  36}'$ is given by
\begin{eqnarray}
  \label{eq:cso}
  \Theta_M{}^\alpha=\Theta_{[AB]}{}^{\!C}{}_{\!D} &\!=\!&
  \delta^C_{[A}\,\theta^{~}_{B]D} \;, \qquad \theta_{AB} = {\rm diag}\{
  \underbrace{1, \dots,1}_p, \underbrace{-1, \dots, -1}_q,
  \underbrace{0, \dots, 0}_r \} \;, 
\end{eqnarray} 
with $A, B = 1, \dots, 8$, and
reproduces the ${\rm CSO}(p,q,r)$ gaugings~\cite{Hull,exhaustive},
where $p+q+r=8$. There are 24 inequivalent gaugings of this type.

Choosing the group ${\rm G}_0={\rm E}_{6(6)}\times {\rm O}(1,1)$,
which is group that can be used to identify gaugings that are related
to compactifications from five dimensions, the branchings of the three
relevant representations are, 
\begin{eqnarray}
{\bf 56} &\rightarrow &  {\bf 1}_{-3}+
\overline{\bf 27}_{-1} +   {\bf 27}_{+1}+{\bf 1}_{+3} \,,\nn\\
{\bf 133} &\rightarrow &  {\bf 27}_{-2}+ {\bf 1}_0+ {\bf 78}_0 +
\overline{\bf 27}_{+2} \,, 
\nn\\
{\bf 912} &\rightarrow &
{\bf 78}_{-3}+ \overline{\bf 27}_{-1}+\overline{{\bf 351}}_{-1}+ {\bf
  351}_{+1}+ {\bf 27}_{+1}+ {\bf 78}_{+3} \,,
\label{Tdec45}
\end{eqnarray}
The first decomposition again captures the split into electric and
magnetic vector fields with the graviphoton transforming in the ${\bf
  1}_{-3}$ and the 27 gauge fields from the five-dimensional theory in
the $\overline{\bf 27}_{-1}$ representation. The couplings between
vector fields and ${\rm E}_{7(7)}$ symmetry generators induced by the
various $\Theta$ components can be summarized in a table analogous
to~(\ref{decblock}),  
\begin{equation}
\begin{tabular}{c|cccc} 
&${\bf 1}_{-3}$ &$\overline{\bf 27}_{-1}$  & ${\bf 27}_{+1}$
&${\bf 1}_{+3}$
\\
\hline
${\bf 27}_{-2}$
&
&${\bf 78} _{-3}$
&$\overline{\bf 351}_{-1}+ \overline{\bf 27} _{-1}$
&${\bf 27}_{+1}$
\\
${\bf 78}_0$
&${\bf 78} _{-3}$
&$\overline{\bf 351}_{-1}+ \overline{\bf 27} _{-1}$
&$ {{\bf 351}}_{+1}+ {\bf {27}} _{+1}$
&${\bf 78} _{+3}$
\\
${\bf 1}_0$
&
& $\overline{\bf 27} _{-1}$
& ${\bf 27}_{+1}$
&
\\
$\overline{\bf 27}_{+2}$
& $\overline{\bf 27} _{-1}$
&$ {{\bf 351}}_{+1}+ {\bf{27}} _{+1}$
&${\bf 78} _{+3}$
&
\\ 
\end{tabular}
\label{decblock-e6}
\end{equation}
The table shows that a gauging involving only electric vector fields
can only live in the ${\bf 78}_{+3}$ representation. Vice versa, every
such embedding tensor automatically satisfies the quadratic 
constraint~(\ref{eq:closure-constraint2}) and thus defines a viable
gauging. These are the theories descending from five dimensions by
Scherk-Schwarz reduction~\cite{sezginvN,AndDauFerrLle,dWST1}.

\subsection{Flux gaugings}
Here we consider gaugings of $N=8$ supergravity that can in principle
be generated by (generalized) toroidal flux compactifications of
type-IIB and M-theory. The proper setting to discuss these theories is
a decomposition \Exc7 group according to ${\rm SL}(2)\times {\rm
  SL}(6)$ and ${\rm SL} (7)$, respectively. For the type-IIB theory
this embedding is realized as 
\begin{equation}
  \label{eq:embed-IIB}
  {\rm E}_{7(7)}~\longrightarrow~
  {\rm SL}(6)\times {\rm SL}(3) ~\longrightarrow~ {\rm SL}(6)\times{\rm
  SL}(2)\times {\rm SO}(1,1) \;. 
\end{equation}
The S-duality group coincides with the ${\rm SL}(2)$ factor. Electric
and magnetic charges transform according to the ${\bf 56}$
representation which branches as
\begin{eqnarray}
  \label{eq:1}
{\bf  56} &\rightarrow& ({\bf 6}', {\bf 1})_{-2} + ({\bf 6}, {\bf
  2})_{-1} 
+ ({\bf 20}, {\bf 1})_0 + ({\bf 6}', {\bf 2})_{+1} +({\bf 6},{\bf
  1})_{+2} \;.\nonumber 
\end{eqnarray}
Here, the $({\bf 6}',{\bf 1})_{-2}$ and $({\bf 6},{\bf 2})_{-1}$
representations descend from graviphotons and two-forms,
respectively, while the four-form generates gauge fields which,
together with their magnetic duals, constitute the $({\bf 20}, {\bf
  1})_0$. The couplings between vector fields and ${\rm E}_{7(7)}$
symmetry generators is summarized in the following
table~\cite{dWST3}, 
\begin{center}
{\scriptsize 
\begin{tabular}{c|ccccc}
&$\!({\bf 6}',{\bf 1})_{-2}\!$ & $\!({\bf 6},{\bf 2})_{-1}\!$ &
$\!({\bf 20},{\bf 1})_{0}\!$ &$\!({\bf 6}',{\bf 2})_{+1}\!$ &
$\!({\bf 6},{\bf 1})_{+2}\!$
\\[.1mm]
\hline
$({\bf 1},{\bf2})_{-3}$ 
& {~} & $({\bf 6},{\bf 1})_{-4}$ & $({\bf 20},{\bf 2})_{-3}$ &
$({\bf 6}',{\bf 3}+{\bf 1})_{-2}$ & $({\bf 6},{\bf 2})_{-1}$
\\[1mm]
$({\bf 15},{\bf 1})_{-2}$ 
& $({\bf 6},{\bf 1})_{-4}$ & $({\bf 20},{\bf 2})_{-3}$ & $({\bf
6}'\!+\!{\bf 84}',{\bf 1})_{-2}$ & $({\bf 6}\!+\!{\bf 84},{\bf
2})_{-1}$ & $({\bf 70}\!+\!{\bf 20},{\bf 1})_{0}$
\\[1mm]
$({\bf 15}',{\bf 2})_{-1}$ 
& $({\bf 20},{\bf 2})_{-3}$ & $({\bf 6}'\!+\!{\bf 84}',{\bf
1})_{-2}+ ({\bf 6}',{\bf 3})_{-2}$ & $({\bf 6}\!+\!{\bf 84},{\bf
2})_{-1}$ & $({\bf 20},{\bf 3}\!+\!{\bf 1})_{0}
   +({\bf 70}',{\bf 1})_0$
& $({\bf 6}'\!+\! {\bf 84}',{\bf 2})_{+1}$
\\[1mm]
$({\bf 1},{\bf 1})_{0}$ 
&$({\bf 6}',{\bf 1})_{-2}$ & $({\bf 6},{\bf 2})_{-1}$ & $({\bf
20},{\bf 1})_{0}$ & $({\bf 6}',{\bf 2})_{+1}$ & $({\bf 6},{\bf
1})_{+2}$
\\[1mm]
$({\bf 35},{\bf 1})_{0}$ 
& $({\bf 6}'\!+\! {\bf 84}',{\bf 1})_{-2}$ & $({\bf 6}\!+\! {\bf
84},{\bf 2})_{-1}$ & $({\bf 70}\!+\!{\bf 70}'\!+\!{\bf 20},{\bf
1})_{0}$ & $({\bf 6}'\!+\! {\bf 84}',{\bf 2})_{+1}$ & $({\bf
6}\!+\! {\bf 84},{\bf 1})_{+2}$
\\[1mm]
$({\bf 1},{\bf 3})_{0}$ 
& $({\bf 6}',{\bf 3})_{-2}$ & $({\bf 6},{\bf 2})_{-1}$ & $({\bf
20},{\bf 3})_{0}$ & $({\bf 6}',{\bf 2})_{+1}$ & $({\bf 6},{\bf
3})_{+2}$
\\[1mm]
$({\bf 15},{\bf 2})_{+1}$ 
& $({\bf 6}\!+\!{\bf 84},{\bf 2})_{-1}$ & $({\bf 20},{\bf
3}\!+\!{\bf 1})_{0}+({\bf 70},{\bf 1})_0$ & $({\bf 6}'\!+\!{\bf
84}',{\bf 2})_{+1}$ & $({\bf 6}\!+\!{\bf 84},{\bf 1})_{+2} + ({\bf
6},{\bf 3})_{+2}$ & $({\bf 20},{\bf 2})_{+3}$
\\[1mm]
$({\bf 15}',{\bf 1})_{+2}$ 
& $({\bf 70}'\!+\!{\bf 20},{\bf 1})_{0}$ & $({\bf 6}'\!+\! {\bf
84}',{\bf 2})_{+1}$ & $({\bf 6}\!+\!{\bf 84} ,{\bf 1})_{+2}$ &
$({\bf 20},{\bf 2})_{+3}$ & $({\bf 6}',{\bf 1})_{+4}$
\\[1mm]
$({\bf 1},{\bf 2})_{+3}$ 
& $({\bf 6}',{\bf 2})_{+1}$ & $({\bf 6},{\bf 3}\!+\!{\bf 1})_{+2}$
& $({\bf 20},{\bf 2})_{+3}$ & $({\bf 6}',{\bf 1})_{+4}$ & {~}
\\
\end{tabular}
}
\end{center}

The entries of the table correspond to the various conjugate
representations of the respective components of the embedding tensor.
Within the ${\bf 912}$ all these components appear with multiplicity~1
apart from the $({\bf 6},{\bf 2})_{-1}$ and $({\bf 6}',{\bf 2})_{+1}$
which appear with multiplicity~2. It follows from the table that an
embedding tensor in the $({\bf 6}',{\bf 1})_{+4}$ defines a purely
electric gauging which thus automatically satisfies the quadratic
constraint. This corresponds to the theory induced by a five-form
flux. A three-form flux on the other hand induces a component of the
embedding tensor in the $({\bf 20,2})_{+3}$ represention, which
involves electric and magnetic gauge fields in the $({\bf
  20,1})_{0}$. Consistency thus requires to further impose the 
quadratic constraint~ (\ref{eq:closure-constraint2}) onto $\Theta$,
leading to~\cite{dWST3}
\begin{equation}
  \label{eq:2}
\varepsilon^{\Lambda\Sigma\Gamma\Omega\Pi\Delta}
\,\theta_{\Lambda\Sigma\Gamma}{}^\sigma\,
\theta_{\Omega\Pi\Delta}{}^\tau =0\;,   
\end{equation}
with $\sigma, \tau = 1, 2$, $\Lambda, \Sigma = 1, \dots, 6$. Here
$\theta_{\Lambda\Sigma\Gamma}{}^\tau$ denotes the components of the
embedding tensor corresponding to the $({\bf 20,2})_{+3}$
representation. The above constraint is precisely the tadpole cancellation condition on the NS-NS and R-R 3-form fluxes.

Gaugings defined by $\Theta$-components with lower ${\rm SO}(1,1)$
grading will correspond to the theories induced by geometric fluxes
(twists), non-geometric compactifications, etc. It follows from the
table that the quadratic constraint~(\ref {eq:closure-constraint2})
leads to more and more consistency conditions among these lower
$\Theta$-components as they tend to stronger mix electric and magnetic
vector fields. It is, however, straightforward to work out these
constraints by branching~(\ref{eq:closure-constraint2}) accordingly
(recall also that the total representation content of this constraint
is given by the ${\bf 133}+{\bf 8645}$) of ${\rm E}_{7(7)}$. Another
representation in the above table which is relevant to string
compactifications is the ${({\bf 84},{\bf 1})}_{+2}$.  It corresponds
to the geometric flux $\tau_{\Lambda\Sigma}{}^\Gamma$ which describes
a ``twisted'' six-torus. The quadratic constraint implies the
following condition, 
\begin{equation}
\tau_{[\Lambda\Sigma}{}^\Gamma\,\tau_{\Pi]\Gamma}{}^\Delta =0\,.
\end{equation}
One may wonder which components of the embedding tensor describe the
non-geometric-fluxes $Q_\Lambda{}^{\Sigma\Gamma}$ and
$R^{\Lambda\Sigma\Gamma}$ obtained from
$\tau_{\Lambda\Sigma}{}^\Gamma$ by applying two subsequent T-dualities
along the directions $\Sigma$ and $\Lambda$, respectively
\cite{Hull:2004in,Shelton:2005cf}. Using the flux/weight
correspondence defined in \cite{fwcorr} we can identify these
non-geometric fluxes with the following representations:
\begin{eqnarray}
Q_\Lambda{}^{\Sigma\Gamma}\in {
({\bf84}^\prime,{\bf2})}_{+1} \quad;\qquad R^{\Lambda\Sigma\Gamma}\in
{({\bf20},{\bf3})}_{0}\,.
\end{eqnarray}
We notice that T-duality changes the ${\rm SL}(2,\mathbb{R})$
representation of the flux on which it acts. We leave a detailed
analysis for future work.

A similar analysis of M-theory fluxes has been performed
in~\cite{D'Auria:2005er}, see also
\cite{Dall'Agata:2005ff,Hull:2006tp}. In this case the relevant
embedding of the torus ${\rm GL}(7)$ is according to
\begin{equation}
  \label{eq:3}
  {\rm E}_{7(7)} ~\longrightarrow~ {\rm SL}(8) ~\longrightarrow~ {\rm
  SL}(7)\times {\rm SO}(1,1) \;.   
\end{equation}
Electric and magnetic charges transform according to the branching
\begin{equation}
{\bf 56}  \rightarrow  {{\bf 7}}'_{-3}+{\bf 21}_{-1}+{{\bf
21}}'_{+1}+ {\bf 7}_{+3} \;,
\end{equation}
where the ${{\bf 7}}'_{-3}$ and the ${\bf 21}_{-1}$ descend from
graviphotons and antisymmetric tensors, respectively. The couplings
between vector fields and ${\rm E}_{7(7)}$ symmetry generators are
given as~\cite{D'Auria:2005er}, 
\begin{center}
\begin{tabular}{c|cccc}
    & ${\bf 7}'_{-3}$ & ${{\bf 21}}_{-1}$ & ${\bf 21}'_{+1}$ & $ {{\bf
  7}}_{+3}$ \\\hline
  ${{\bf 7}}_{-4}$ & ${\bf 1}_{-7}$ & ${{\bf 35}}_{-5}$ & $({\bf
  140}'+{\bf 7}')_{-3}$ & $({{\bf 28}}+{{\bf 21}})_{-1}$ \\
  ${\bf 35}'_{-2}$ & ${{\bf 35}}_{-5}$ &${\bf 140}'_{-3}$ & $({{\bf
  21}}+{{\bf 224}})_{-1}$ &$({{\bf 21}}'+{{\bf 224}}')_{+1}$ \\
  ${\bf 48}_0$ & $({\bf 140}'+{\bf 7}')_{-3}$ & $({{\bf 21}}+{{\bf
  28}}+{{\bf 224}})_{-1}$ &
   $({{\bf 21}}'+{{\bf 28}}'+{{\bf 224}}')_{+1}$
  & $({{\bf 140}}+{{\bf 7}})_{+3}$\\
  ${\bf 1}_0$ & ${\bf 7}'_{-3}$ & ${{\bf 21}}_{-1}$ & ${\bf 21}'_{+1}
  $ & $ {{\bf 7}}_{+3}$ \\
  ${{\bf 35}}_{+2}$ & $({{\bf 21}}+{{\bf 224}})_{-1}$ & $({{\bf
  21}}'+{{\bf 224}}')_{+1}$ &
  ${{\bf 140}}_{+3}$ & ${\bf 35}'_{+5}$ \\
  ${\bf 7}'_{+4}$ & $({{\bf 28}}'+{{\bf 21}}')_{+1}$ & $({{\bf 140}}+
  {{\bf 7}})_{+3}$ & ${\bf 35}'_{+5}$ & ${\bf 1}_{+7}$ \\
\end{tabular}
\end{center}
The table shows that an embedding tensor in the ${\bf 1}_{+7}$ and in
the ${{\bf 35}}'_{+5}$ representation define electric gaugings that
automatically satisfy the quadratic constraint. They describe the
theories obtained by switching on in eleven dimensions a seven-form
$g_7$ and a four-form flux $g_{\Lambda\Sigma\Gamma\Delta}$,
respectively. The former theory has in fact already been considered
in~\cite {Aurilia:1980xj}. An embedding tensor in the ${\bf 140}_{+3}$
corresponds to the parameters $\tau_{\Lambda\Sigma}{}^{\Xi}$ of a
geometric flux and is subject to the quadratic
constraint~(\ref{eq:closure-constraint2}), 
\begin{equation}
  \label{eq:4}
  \tau_{[\Lambda\Sigma}{}^{\Omega} \tau_{\Gamma]\Omega}{}^{\Xi}= 0\;,   
\end{equation}
with $\Lambda, \Sigma = 1, \dots, 7$, corresponding to the Jacobi
identity of the associated gauge algebra. If $g_7\,,
g_{\Lambda\Sigma\Gamma\Delta}$ and the geometric flux
$\tau_{\Lambda\Sigma}{}^{\Gamma}$ are switched on together, the second
order constraint on the embedding tensor, as was shown in
\cite{D'Auria:2005er}, yields the additional condition
\begin{equation}
\tau_{[\Lambda\Sigma}{}^\Delta\,g_{\Gamma\Pi\Omega]\Delta} =0\,,
\end{equation}
originally found in \cite{Dall'Agata:2005ff}.

\subsection{Gaugings of six-dimensional origin}
In this subsection we demonstrate our method for gaugings that
arise, in particular, from a two-fold Scherk-Schwarz reduction from six
space-time dimensions \cite{ScherkSchwarz,CSS}. The Scherk-Schwarz
reduction of maximal supergravity from $D=5$ to $D=4$ spacetime
dimensions was first constructed in \cite{sezginvN} and recently this
theory was obtained more directly in four spacetime dimensions by
gauging \cite{AndDauFerrLle}. For a general treatment of
Scherk--Schwarz reductions in relation to gauged maximal
supergravities, see \cite{dWST1}, where the Scherk-Schwarz reduced
maximal supergravity from $D=6$ to $D=5$ was constructed as a gauging of
five-dimensional supergravity.

The proper choice for ${\rm G}_{0}$ is the maximal subgroup
\begin{equation}
  \label{eq:5}
  {\rm
  E}_{7(7)} ~\longrightarrow~ {\rm SO}(5,5)\times {\rm SL}
  (2,\mathbb{R})\times{\rm O}(1,1) \;,  
\end{equation}
where ${\rm SO}(5,5)$ represents the non-linear symmetry group of
maximal supergravity in six dimensions and ${\rm
  SL}(2,\mathbb{R})\times{\rm O}(1,1)$ is the group corresponding to
the reduction on a two-torus.  Electric and magnetic charges branch as
\begin{equation}
  \label{eq:6}
{\bf 56}\rightarrow {({\bf1},{\bf2})}_{-2}+
  {(\overline{{\bf16}},{\bf1})}_{-1}+{({\bf10},{\bf2})}_{0} +  
  ({\bf16},{\bf1})_{+1}+({\bf1},{\bf2})_{+2} \;,  
\end{equation}
where the ${({\bf1},{\bf2})}_{-2}$ and
${(\overline{{\bf16}},{\bf1})}_{-1}$ correspond to graviphotons and
six-dimensional vectors, respectively, while the
${({\bf10},{\bf2})}_{0}$ combines the electric vectors and their
magnetic duals descending from the self-dual two-forms in
six-dimensions.  Their couplings to the ${\rm E}_{7(7)}$ symmetry
generators are summarized in the table below,
\begin{center}
{\scriptsize \begin{tabular}{c|ccccc}
   & ${({\bf1},{\bf2})}_{-2}$ & ${(\overline{{\bf16}},{\bf1})}_{-1}$ & ${
   ({\bf10},{\bf2})}_{0}$ &
   ${({\bf16},{\bf1})}_{+1}$ & ${({\bf1},{\bf2})}_{+2}$ \\ \hline
   ${({\bf10},{\bf1})}_{-2}$ &  &  ${({\bf16},{\bf1})}_{-3}$ &
   ${({\bf1}+{\bf45},{\bf2})}_{-2}$ 
   & ${(\overline{{\bf16}}+{\bf144},{\bf1})}_{-1}$&  ${
   ({\bf10},{\bf2})}_{0}$ \\
   ${({\bf16},{\bf2})}_{-1}$& ${({\bf16},{\bf1})}_{-3}$ &
   ${({\bf1}+{\bf45},{\bf2})}_{-2}$ 
   & ${(\overline{{\bf16}},{\bf3}+{\bf1})}_{-1}
   +{({\bf144},{\bf1})}_{-1}$ & ${({\bf10}+{\bf120},{\bf2})}_{0}$ &
   ${({\bf16},{\bf3}+{\bf1})}_{+1}$ \\ 
   ${({\bf1},{\bf3})}_{0}$ & ${({\bf1},{\bf2})}_{-2}$ &
   ${(\overline{{\bf16}},{\bf3})}_{-1}$ & 
   ${({\bf10},{\bf2})}_{0}$ & ${({\bf16},{\bf3})}_{+1}$ &
   ${({\bf1},{\bf2})}_{+2}$\\ 
   ${({\bf1},{\bf1})}_{0}$ & ${({\bf1},{\bf2})}_{-2}$ &
   ${(\overline{{\bf16}},{\bf1})}_{-1}$ & 
   ${({\bf10},{\bf2})}_{0}$ &  ${({\bf16},{\bf1})}_{+1}$ &
   ${({\bf1},{\bf2})}_{+2}$ \\ 
   ${({\bf45},{\bf1})}_{0}$ & ${({\bf45},{\bf2})}_{-2}$ &
   ${(\overline{{\bf16}}+{\bf144},{\bf1})}_{-1}$
   & ${({\bf10}+{\bf120},{\bf2})}_{0}$ &
   ${({\bf16}+\overline{{\bf144}},{\bf1})}_{+1}$ & 
   ${({\bf45},{\bf2})}_{+2}$ \\
   ${(\overline{{\bf16}},{\bf2})}_{+1}$
  &${(\overline{{\bf16}},{\bf3}+{\bf1})}_{-1}$ & 
  ${({\bf10}+{\bf120},{\bf2})}_{0}$ & ${({\bf16},{\bf3}+{\bf1})}_{+1}
  +{(\overline{{\bf144}},{\bf1})}_{+1}$
  &${({\bf1}+{\bf45},{\bf2})}_{+2}$ &
  ${(\overline{{\bf16}},{\bf1})}_{+3}$ \\ 
  ${({\bf10},{\bf1})}_{+2}$ &  ${({\bf10},{\bf2})}_{0}$
  &${({\bf16}+\overline{{\bf144}},{\bf1})}_{+1}$ 
  & ${({\bf1}+{\bf45},{\bf2})}_{+2}$ & ${
   (\overline{{\bf16}},{\bf1})}_{+3}$ &  \\
\end{tabular}}\label{Tso55}
\end{center}
This shows that an embedding tensor in the
${(\overline{{\bf16}},{\bf1})}_{+3}$ defines a consistent electric
gauging corresponding to the theory obtained by giving a $T^{2}$ flux
to the six-dimensional vector field strength.  A Scherk-Schwarz
gauging is defined by an embedding tensor in the $({\bf 45},{\bf
  2})_{+2}$, i.e.\ a tensor of type $\theta_{u,mn}$ with $m, n =
1,\ldots, 10$, $u=1,2$.  This corresponds to identifying the two gauge
group generators $X_{u}=\theta_{u,mn}\,t^{mn}$ generating a subgroup
of ${\rm SO}(5,5)$ associated with the dependence of the fields on the
internal $T^2$ according to the Scherk-Schwarz ansatz, which couple to
the two graviphotons.  As this gauging a priori involves electric and
magnetic vector fields, the quadratic
constraint~(\ref{eq:closure-constraint2}) poses a nontrivial
restriction,
\begin{equation}
  \label{2ndord}
\epsilon^{uw}\,\eta^{mn}\,\theta_{u,mp}\,\theta_{w,nq}=0\,,
\end{equation}
which implies $[X_u,X_v]=0$. This is consistent as these generators
must commute in the multiple Scherk-Schwarz reduction.  The complete
gauge algebra in four dimensions takes the form
\begin{eqnarray}
\left[X_u,\,X_v\right]&=&0\;,\qquad
\left[X_u,\,X_\sigma\right]~\propto~
\theta_{u,mn}
(\Gamma^{mn})_\sigma{}^\tau\,X_\tau\;,\nonumber\\[1ex]
\left[X_u,\,X_{mw}\right]&\propto&
\theta_{u,mn}\,\eta^{np}\,X_{pw}\;,\qquad
\left[X_\sigma,\,X_\tau\right]~\propto~
\epsilon^{uv}\theta_{u,mn}
(\Gamma^{mnp})_{\sigma\tau}\,X_{pv}\;, 
\end{eqnarray}
with ${\rm SO}(5,5)$ $\Gamma$-matrices $\Gamma^{m}_{\sigma\tau}$.
We intend to give a detailed analysis of this theory elsewhere.

\vspace{8mm}

\noindent
{\bf Acknowledgement}\\
\noindent
We are grateful to Mathijs de Vroome, Hermann Nicolai, and Martin
Weidner for enjoyable and helpful discussions. B.d.W. thanks the
Galileo Galilei Institute for Theoretical Physics for its hospitality
and INFN for partial support during the completion of this work. The
work is partly supported by EU contracts MRTN-CT-2004-005104 and
MRTN-CT-2004-512194, by INTAS contract 03-51-6346 and by NWO grant
047017015. \bigskip

%

%
\end{document}